\documentclass[aps,prb, twocolumn, nofootinbib, superscript address]{revtex4-2}
\usepackage{amsmath}
\usepackage{amssymb}
\usepackage{graphicx}
\usepackage{dcolumn}    
\usepackage{bm}     
\usepackage{epstopdf}
\usepackage{graphicx}
\usepackage{tikz}
\usepackage{lipsum}
\usepackage{pgf}
\usepackage{float}
\usepackage{hyperref}
\usepackage{tabularx}
\usepackage{subfigure}
\usepackage{float}

\newcommand {\pd} [2] {\frac{\partial #1}{\partial #2}}
\newcommand {\td} [2] {\frac{d #1}{d #2}}

 \newcommand {\beq}{\begin{equation}}
\newcommand {\eeq}{\end{equation}}
\newcommand {\bea}{\begin{eqnarray}}
\newcommand {\eea}{\end{eqnarray}}

\begin{document}
\title{Spin-Hall effect due to the bulk states of topological insulators: Extrinsic contribution to the conserved spin current}

\author{James H. Cullen}
\affiliation{School of Physics, The University of New South Wales, Sydney 2052, Australia}
\author{Dimitrie Culcer}
\affiliation{School of Physics, The University of New South Wales, Sydney 2052, Australia}
\date{\today}
\begin{abstract}
The substantial amount of recent research into spin torques has been accompanied by a revival of interest in the spin-Hall effect. This effect contributes to the spin torque in many materials, including topological insulator/ferromagnet devices, Weyl semimetals, and van der Waals heterostructures. In general the relative sizes of competing spin torque mechanisms remain poorly understood. Whereas a consensus is beginning to emerge on the evaluation of a conserved spin current, the role of extrinsic disorder mechanisms in the spin-Hall effect has not been clarified. In this work we present a comprehensive calculation of the \textit{extrinsic} spin Hall effect while focussing on the bulk states of topological insulators as a prototype system and employing a fully quantum mechanical formalism to calculate the proper spin current. Our calculation of the proper spin current employs a $4\times4$ $\boldsymbol{k}\cdot\boldsymbol{p}$ Hamiltonian describing the bulk states of topological insulators. At the same time, we provide a qualitative explanation of the proper spin currents calculated based on an effective $2\times2$ Hamiltonian obtained via a Schrieffer-Wolf transformation. We find that the extrinsic contribution to the proper spin current, driven by side jump, skew scattering and related mechanisms, is of a comparable magnitude to the intrinsic contribution, making it vital to take such disorder effects into account when seeking to understand experiments. Among the scattering effects considered, side jump scattering is the primary contributor to the extrinsic spin Hall effect. The total spin susceptibility calculated here is too small to explain experimentally measured spin torques, hence we expect the spin Hall effect to make a negligible contribution to the spin torque in topological insulator structures.
\end{abstract}
\maketitle 

\section{Introduction}
The spin Hall effect (SHE), the generation of a transverse spin current in response to an applied electric field, has witnessed a surge of renewed interest in recent years due to its relevance to spin torques, which provide a promising avenue towards electrical control of magnetic degrees of freedom\cite{Nikolic_SOT, Manchon2019, brataas2012, wu2021,Shao2021RoadmapOS,Ramaswamy2018,Manchon2019}. Following its prediction, \cite{DYAKONOV1971, DYAKONOV1971459, SHE-Hirsch-PRL-1999, Shuichi-Science-2003, Sinova-Dimi-PRL-2004} the SHE has been observed in semiconductors \cite{Exp-SHE-PRL-2005,Exp-SHE-Science-2004} and metals.\cite{Bi-SHE-APL,ESHE-Copper-PRL-2011,Intrinsic-SH-metal-2011,Gaint-SHE-Ta-Sci,Giant-SHE-SOT-APL-2012,Giant-SHE-CuBi-PRL-2012,Hoffmann-PRL,SHE-Metal-APL-2014,SHE-Exp-Pt-ACSnano-2022,Giant-SHE-Metal-Adv-Mat-2022} Recently, it has been studied in more exotic materials such as topological insulators, Weyl semimetals\cite{muduli2019evaluation,zhao2020observation} and van der Waals heterostructures.\cite{avsar2014spin,benitez2020tunable,safeer2019room,safeer2020spin,DASTGEER2022,herling2020gate,ghiasi2019charge,hoque2022charge} Spin torque devices that utilise the spin Hall effect do this by generating spin currents in a material with spin-orbit coupling which flow into an adjacent magnetic material in which the polarised spins exert a torque on the magnetization. While spin currents of intrinsic origin have received most theoretical attention, \cite{Akzyanov20182, Ghosh2018, liu2023, Cong-CC-PRB} the need persists for a more profound understanding of extrinsic spin currents, which form the subject of this work.

The difficulty in theoretically studying the spin Hall effect lies in the definition of the spin current. The intuitive and conventional definition of the spin current is the product of the spin and velocity operators.\cite{Toplogical-SC-PRB-Schmeltzer,Spin-Hall-inhomo-E,Mingche-PRB-SC-2006, Conventional-SC-Sci-Rep,PhysRevB.100.245430,PhysRevB.97.195127,PhysRevLett.114.107201,PhysRevLett.117.146403,PhysRevB.71.245327,PhysRevB.74.085315} However, the generation of a spin current generally requires spin-orbit coupling, which causes spin precession and hence non-conservation. This makes the conventional definition meaningless in most contexts of interest. One way to address this is to circumvent the spin current altogether by calculating directly the spin density and/or spin accumulation.\cite{Tatara-PRB-2018,Tatara-PRB-Letter,Spin-toroidization-XiaoDi-PRB-2018,Spin-M-Quadrupole-prb-Yanase-PRB-2019} However, there are many systems where the spin current itself is the quantity of interest, including magnetic systems with sizable spin-Hall torques, discussed below. To evaluate the spin Hall effect in spin-orbit coupled systems one needs to evaluate the proper spin current, which takes into account the torque dipole arising from spin precession.\cite{Dimi-PRL-2004, SC-Shuichi-PRB-2004, Anisotropic-SHE-PRL-2010, PhysRevLett.109.246604, Conserved-SC-Mott-PRB-Cong-2018, Defintion-SC-PRL-2006-Qian,CSC-2D-hole-Qian-PRB-2008,SHE-insulator-PRB-2020,Cong-CC-PRB} The torque dipole is notoriously difficult to evaluate for Bloch electrons, and, until recently, available theories only provided results for simple 2D effective spin-1/2 models, with the spin primarily in the plane. However, there have been new developments in the understanding of the SHE and the proper spin current. Two recent theories have brought to light the relationship between the intrinsic proper spin current and the underlying topological structure of the Hilbert space, with very similar results.\cite{Cong-CC-PRB,liu2023} A quantum mechanical study determined the intrinsic contribution to the proper spin current and SHE, relating the intrinsic proper spin current to the inter-band matrix elements of the Berry connection.\cite{liu2023} The results are broadly in agreement with the evolving semiclassical understanding of the SHE,\cite{Cong-CC-PRB} in which the intrinsic proper spin current is expressed in terms of the Berry curvature. The final expressions in the two studies only differ in the position of the spin matrix elements.

A complete description of the spin Hall effect must necessarily include extrinsic mechanisms due to impurity scattering. Scattering introduces sizeable transport effects that are independent of the disorder strength making them indistinguishable from intrinsic mechanisms. This has been studied in depth using the conventional definition of the spin current, \cite{sugimoto2006, hankiewicz2006, inoue2004suppression, raimondi2005, mishchenko2004spin, liu2006vanishing} and is also known to occur in the anomalous Hall effect. \cite{sinitsyn2007,sinitsyn2007semiclassical,nagaosa2010anomalous} Extrinsic effects on the proper spin current have been studied in the past in Ref. \onlinecite{sugimoto2006} which presented a formula for the proper spin current that included disorder effects. This work clearly showed that the inclusion of disorder is vital for an accurate calculation of the proper spin current, as it showed that disorder effects can sometimes be the dominant contribution to the proper spin current. However, the complex approach of Ref. \onlinecite{sugimoto2006}, which involves fictitious electric and magnetic fields, does not directly relate to topological quantities and is, furthermore, restricted to specific 2D spin-1/2 systems with the spin lying in the plane. This calls for a general, systematic theory of disorder in the context of the proper spin current and the SHE. 


\begin{figure}[t!]
	\centering
	\includegraphics[width = \columnwidth]{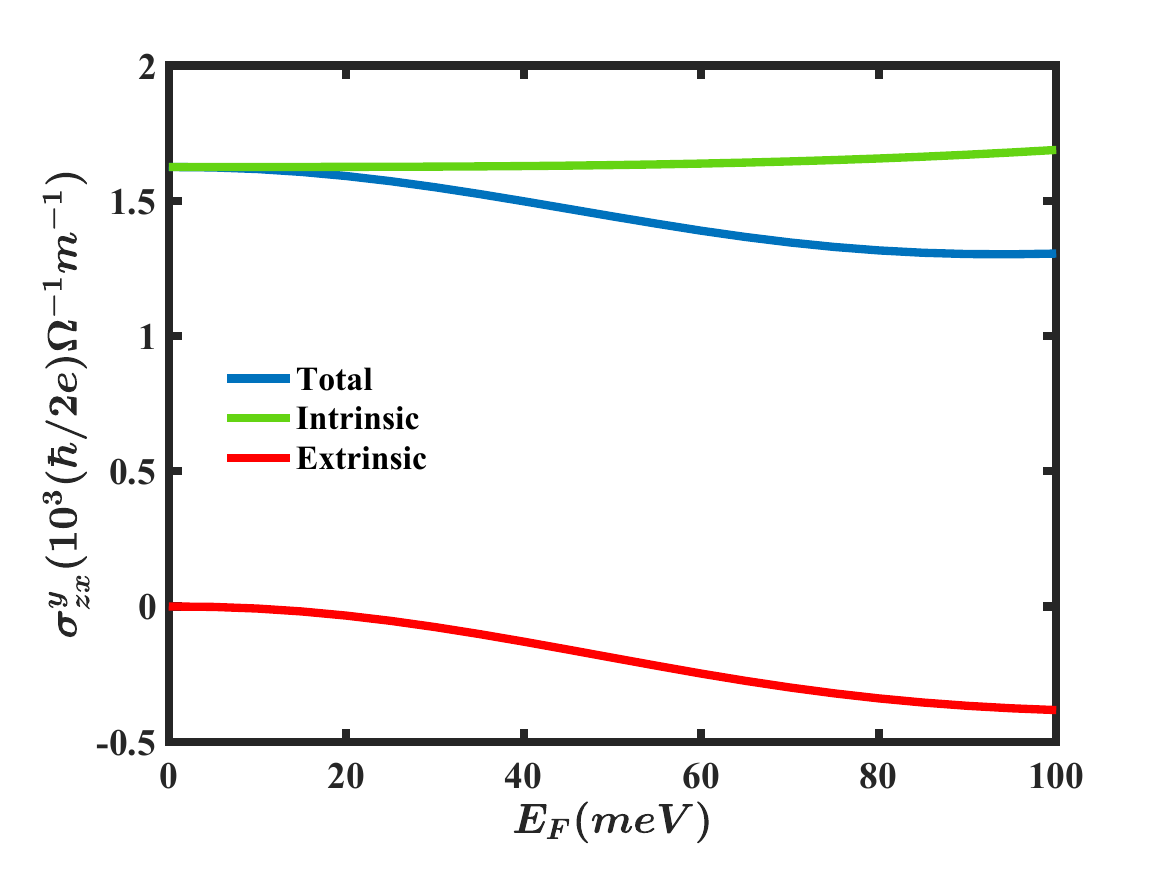}
	\caption{\label{Fig1} The spin Hall conductivity $\sigma^{y}_{zx}$ vs the Fermi energy $E_F$ for the TI bulk states in Bi$_2$Se$_3$ with Zeeman field $\boldsymbol{m}\parallel \hat{z}$ and $|m|=10$ $\mu$eV (Bi$_2$Se$_3$ parameters from Ref.~\onlinecite{Liu2010}).}
\end{figure}

In light of the above, in this paper we develop a fully quantum mechanical formalism for the calculation of proper spin currents of extrinsic origin, including skew scattering and side jump. This work aims to (i) provide a general blueprint for calculating the full spin current in the presence of disorder, and (ii) apply this method to calculate the full spin Hall effect due to the bulk states of 3D topological insulators, focussing on the disorder contributions. We determine the extrinsic spin currents up to zeroth order in the scattering time $\tau$, which we take to be a measure of the disorder strength. We consider the effects of a scalar disorder potential combined with band structure spin-orbit coupling, leading to skew scattering and side jump, with the latter incorporating an electric field correction to the scattering term. \cite{atencia2022, culcer_sj_2010, culcer2010, Bi2013} The spin current contributions from these two mechanisms appear to zeroth order in the scattering time. They are independent of disorder strength and appear due to the disorder-independent part of the non equilibrium density matrix $\rho^{(0)}_E$. Hence, these extrinsic contributions to the spin current compete with the intrinsic contributions,\cite{culcer_sj_2010, Bi2013} which by definition are independent of the disorder strength. It is crucial to consider disorder effects on the proper spin current: not only is this the only physically meaningful definition, but, as the study of the intrinsic case shows, many of the conventional spin current terms exactly are canceled by the torque dipole correction,\cite{liu2023} and it is natural to expect similar cancellations in the extrinsic contributions. 

We consider, as a prototype system, the bulk states of topological insulators (TI). This choice is motivated by the observation that topological insulators are excellent candidates for building spin torque devices due to their high charge to spin conversion efficiency. Spin torques are especially strong in topological insulators \cite{lu2022, mogi2021, che2020, Li2019, Li2014, liu2018, Zhang2018, Rodriguez-Vega2017, Song2018, Shi2018, bonell2020, wu2019, Liu2021} and large spin torques have been demonstrated experimentally in a plethora of ferromagnet(ferrimagnet)/TI heterostructures,\cite{Mellnik2014, Wang2015, Jamali2015, Kondou2016, Fanchiang2018, zhu2021, Ramaswamy2019, Baker2015, Deorani2014, Shiomi2014, Wang2016, singh2020, zheng2020, Fan2022, Fan2014, fan2016} including room-temperature magnetization switching.\cite{Wang2017, Han2017, Khang2018Acon, Dc2018} The extent to which the spin Hall effect contributes to the spin torque is yet to be conclusively settled, \cite{Jamali2015, Wang2016, gao2019, mogi2021} and a full account of the spin-Hall effect cannot be given without considering the extrinsic contribution to the physical spin current.

Hence, for concreteness, after introducing the general formalism for calculating the proper spin current in the presence of disorder, we determine the linear response of the TI bulk density matrix to an electric field in a system with short ranged non magnetic impurities. We then formulate an expression for the extrinsic proper spin current using the same approach that was used in Ref. \onlinecite{liu2023} for the intrinsic case. The main results we present are: (i) The extrinsic spin Hall effect in TIs is of a similar magnitude to the intrinsic spin Hall effect, as shown in Fig. \ref{Fig1}, which may be regarded as a summary of the central results of this work; (ii) The largest component of the extrinsic spin current, primarily driven by side jump scattering, should generate a field-like spin torque; (iii) The size of the spin currents generated by the spin Hall effect in TIs should have a negligible contribution on the total spin torque. As Fig. \ref{Fig1} shows, the spin conductivities due to the SHE are of the order $10^3(\hbar/2e)\Omega^{-1}m^{-1}$, which is 1-2 orders of magnitude smaller than the spin conductivities reported in experiment.\cite{Mellnik2014,Han2017,Dc2018,Wang2017} Hence, the only potentially sizeable bulk state contribution to the TI spin torque is the spin transfer torque, introduced in Ref.~\onlinecite{cullen2023}. We note, at the same time, that the expression we present for the spin current is general and can be applied in further studies to other materials of interest.

This paper is organised as follows: first in sections \uppercase\expandafter{\romannumeral2\relax} and \uppercase\expandafter{\romannumeral3\relax} we present the model Hamiltonian and linear response formalism based on the density matrix. Next we discuss the calculation of the proper spin current and present a general formula for its evaluation. Then in section \uppercase\expandafter{\romannumeral4\relax} we present our results for the extrinsic spin conductivity in topological insulators, focusing on Bi$_2$Se$_3$ for concreteness. We show that different components of the spin current have different dependencies on the impurity strength and Zeeman field. In section \uppercase\expandafter{\romannumeral5\relax} we discuss the role of the extrinsic and intrinsic spin Hall effect in topological insulator spin torques and potential ways to measure the extrinsic spin Hall effect. Lastly, we discuss the applicability of our proper spin current calculation to other systems.

\section{Model Hamiltonian}
Bulk TI states are described by the Hamiltonian $H_0 = \varepsilon_{\mathbf{k}}+ H_{so} + U + e\boldsymbol{E}\cdot\boldsymbol{r}$, where $\varepsilon_{\mathbf{k}} = C_{0}+C_{1} k_{z}^{2}+C_{2} k_{\parallel}^{2}$, the spin-orbit Hamiltonian $H_{so}$ is given by Ref. \onlinecite{Liu2010}, U is the disorder contribution and $e\boldsymbol{E}\cdot\boldsymbol{r}$ is the electric potential. The disorder contribution is calculated using the Born approximation and the electric potential is treated perturbatively. In the basis $\{ \frac{1}{2},-\frac{1}{2},\frac{1}{2},-\frac{1}{2} \}$ the spin-orbit Hamiltonian is:
\begin{equation}
	\begin{aligned}
		H_{so} & = &
		\begin{pmatrix}
			-\mathcal{M} + m_z & m_- & \mathcal{B} k_{z} & \mathcal{A}k_{-} \\
			m_+ & -\mathcal{M} - m_z & \mathcal{A} k_{+} & -\mathcal{B}k_z \\
			\mathcal{B}k_z & \mathcal{A} k_{-} & \mathcal{M} + m_z & m_- \\
			\mathcal{A} k_{+} & -\mathcal{B} k_{z} & m_+ &  \mathcal{M} - m_z
		\end{pmatrix}, 
	\end{aligned}
\end{equation}
with $\mathcal{M} = M_{0}+M_{1} k_{z}^{2}+M_{2} k_{\parallel}^{2}$, $\mathcal{A} = A_{0}+A_{2} k_{\parallel}^{2}$, $\mathcal{B} = B_{0}+B_{2} k_{z}^{2}$, $k_{\parallel}^{2} = k_{x}^{2}+k_{y}^{2}$, $k_{\pm} = k_{x} \pm i k_{y}$ and $m_{\pm}=m_x \pm i m_y$. In our Hamiltonian we include a small Zeeman field $\boldsymbol{m}$ to remove spin degeneracy, for most of our calculations we set it to be $10$ $\mu$eV. It can be thought of as spin splitting due to an applied external magnetic field, as one is often used in spin torque experiments.\cite{Ramaswamy2018} For most of the calculation we have ignored the hexagonal warping terms due to the added complexity. However, their effects have been calculated and are discussed in the results section.

This model is only accurate near the band center and is valid in the regime $k<4\times10^{8}$ m$^{-1}$. We use this $\boldsymbol{k}\cdot\boldsymbol{p}$ Hamiltonian here despite its limitations as it allows us to apply our transport formalism to the problem. Our approach has an advantage over other numerical models and methods that struggle to properly treat disorder.

\section{Linear response}

We use a kinetic equation formalism to calculate the linear response of the bulk states to an electric field $\boldsymbol{E}$, starting from the quantum Liouville equation as described in Refs.~\cite{Culcer2017,Bi2013,atencia2022}. This transport formalism can be thought of as the quantum analog of the Boltzmann equation. The linear response of the bulk states is characterised by following kinetic equation
\begin{equation}\label{KE1}
    \frac{\partial \langle\rho_E\rangle}{\partial t} + \frac{i}{\hbar}\, [H_0, \langle\rho_E\rangle] + \hat J_0 (\langle\rho_E\rangle) = \frac{e\boldsymbol{E}}{\hbar}\cdot \frac{D\langle\rho_0\rangle}{D\boldsymbol{k}}\,,
\end{equation}
where $\langle\rho_0\rangle$ is the equilibrium density matrix, $\langle\rho_E\rangle$ is the non-equilibrium density matrix to first order in the electric field and, $\hat{J}$ contains the disorder contribution. The equilibrium density matrix is simply the Fermi-Dirac distribution. Here $\langle \rho \rangle$ represents the disorder averaged density matrix.

To solve this kinetic equation we break the density matrix $\langle \rho_E\rangle$ up into two components; $n_E$ a band diagonal part and $S_E$ a band off-diagonal part. In the steady state limit the kinetic equation for diagonal part simplifies greatly, and the solution can be found by solving the equation 
\begin{equation}\label{DiagTau}
    \left[\hat J_0 (n_E)\right]_{nn} = \frac{e\boldsymbol{E}}{\hbar}\cdot \frac{\partial f^n_k}{\partial\boldsymbol{k}}\,.
\end{equation} 
Carrying out the time integral for the off-diagonal part gives
\begin{equation}\label{OffD}
    S_{E,nm}=-i\hbar\frac{e\boldsymbol{E}\cdot\mathcal{R}_{nm}(f^n_k-f^m_k)-\left[\hat{J}_{0}(n_E)\right]_{nm}}{\epsilon^n_k-\epsilon^m_k}\,,
\end{equation}
where, $\mathcal{R}$ is the Berry connection and $\epsilon^n_k$ is the energy of the eigenstate in band $n$ with wavevector $k$. The first part of (\ref{OffD}) is purely intrinsic and will be ignored in this calculation, the spin current due to this term was studied in Ref. \onlinecite{liu2023}. The disorder contribution $\hat{J}$ is calculated in the Born approximation. The Born approximation scattering term is
\begin{equation}\label{Born}
    J(\hat{f})=\frac{1}{\hbar^2}\int_{0}^{\infty} dt^{\prime} \langle e^{-\eta t^{\prime}} \big[\hat{U},e^{-i\hat{H}t^{\prime}/\hbar}\big[\hat{U},\hat{f}\big] e^{i\hat{H}t^{\prime}/\hbar}\big]\rangle_{\boldsymbol{kk}}\,.
\end{equation}
Here we consider short ranged scalar disorder of the form $U_i=U_0\delta(r_i-r)$. The way in which this scattering integral is calculated is outlined in Ref. \onlinecite{Culcer2017}. Calculating the scattering integral (\ref{Born}) and solving (\ref{DiagTau}) will give the band diagonal response to order $-1$ in the impurity density. Substituting this solution into (\ref{OffD}) will give the extrinsic off-diagonal density matrix to zeroth order in the impurity density.

To find the diagonal part of the non-equilibrium density matrix to zeroth order in the impurity density we need included some extra corrections to the scattering term. The band-diagonal kinetic equation to the zeroth order in the impurity density is
\begin{equation}\label{KE2}
    \frac{\partial n_E^0}{\partial t} + \frac{i}{\hbar}\, [H_0, n_E^0] + \hat J_0 (n_E^0) = -\hat J_E(\langle\rho_0\rangle)-\hat J_{sk}(n_E^{-1})\,,
\end{equation}
where $\hat J_E$ is a electric field correction to the scattering term\cite{atencia2022} and $\hat J_{sk}$ is the scattering term found by substituting (\ref{OffD}) back into the band-diagonal part of the scattering integral (\ref{Born}). In semi-classical calculations $\hat J_E$ is considered to be part of side-jump scattering and $\hat J_{sk}$ part of skew-scattering.\cite{atencia2022,culcer2010,Bi2013} We find that these scattering terms are crucial for a proper calculation of the extrinsic spin current.

The general definition of the conserved spin current is $\hat{\mathcal{J}}^i_j = \td{}{t} \, (\hat{r}_j\hat{s}_i)$. The regularly used conventional spin current $J^i_j=\frac{1}{2}\{ s_i, v_j \}$ fails to account for the absence of spin conservation in materials with spin orbit coupling. As has been shown in a recent paper the proper intrinsic spin current can be captured by the following equation\cite{liu2023}
\begin{equation}
    \mathcal{J}^i_j = \sum_k \left[\frac{e \boldsymbol{E}}{\hbar}\,\times\,\sum_m\boldsymbol{\Sigma}^i_m f_m\right]_{j}.
\end{equation}
Where $\boldsymbol{\Sigma}$ is a topological quantity related to the Berry connection and spin operator. This formula only captures intrinsic spin currents and does not account for disorder contributions. However, disorder contributions can be straightforwardly calculated using the same methodology used in Ref. \onlinecite{liu2023}. Here we will evaluate these extra contributions.

The proper spin current can be broken up into two parts: (i) The conventional spin current $\{ s_i, v_j \}$, the extrinsic terms due to the conventional spin current are
\begin{equation}\label{SCcon}
    J^i_j = \sum_{m,\boldsymbol{k}} \frac{s^i_{mm} n^E_m}{\hbar} \pd{\varepsilon_m}{k_j} -\frac{s^i_{mm}}{2}\{\mathcal{R}_j,[\hat{J}_{0}(n_E)]_{od}\}_{mm}\,.
\end{equation}
(ii) The torque dipole correction $\{\frac{\partial s_i}{\partial t},\hat{r}_j\}$. 

In order to evaluate the torque dipole in the proper spin current we must allow the density matrix to have terms that are off-diagonal in the wave vector $k$. To do this we expand the density matrix $\rho_{\boldsymbol{kk}^{\prime}}$ perturbatively in terms of a small off-diagonal wavevector $\boldsymbol{Q}$, such that ${\bm k} \equiv {\bm q}_+ = {\bm q} + {\bm Q}/2$ and ${\bm k}' \equiv {\bm q}_- = {\bm q} - {\bm Q}/2$. Using this transformation we can reformulate our kinetic equation. This will give successive equations each of increasing order in the perturbation $\boldsymbol{Q}$, the zeroth order equation is simply (\ref{KE1}). The kinetic equation to first order in $\boldsymbol{Q}$ is
\begin{equation}\label{QKE}
    \displaystyle \pd{\rho_{{\bm q}{\bm Q}}}{t} + \frac{i}{\hbar} \, [H_{0{\bm q}}, \rho_{{\bm q}{\bm Q}}] + \hat J_0(\rho_{{\bm q} {\bm Q}}) = \displaystyle - \frac{i{\bm Q}}{2\hbar} \cdot \bigg\{\frac{DH_{0{\bm q}}}{D{\bm q}}, \rho_{\bm q} \bigg\}\,.
\end{equation}
Where $\rho_{\boldsymbol{q}}$ is simply the solution to (\ref{KE1}) and $\rho_{\boldsymbol{q} \boldsymbol{Q}}$ is the density matrix to first order in $\boldsymbol{Q}$. This kinetic equation is solved in an identical manner to (\ref{KE1}), as the scattering term to linear order in $\boldsymbol{Q}$ is identical to the scattering term diagonal in wavevector. Due to the form of the torque dipole operator only solutions to the first order in $\boldsymbol{Q}$ are required for the proper spin current calculation. 

\begin{figure}[t!]
	\centering
	\includegraphics[width = \columnwidth]{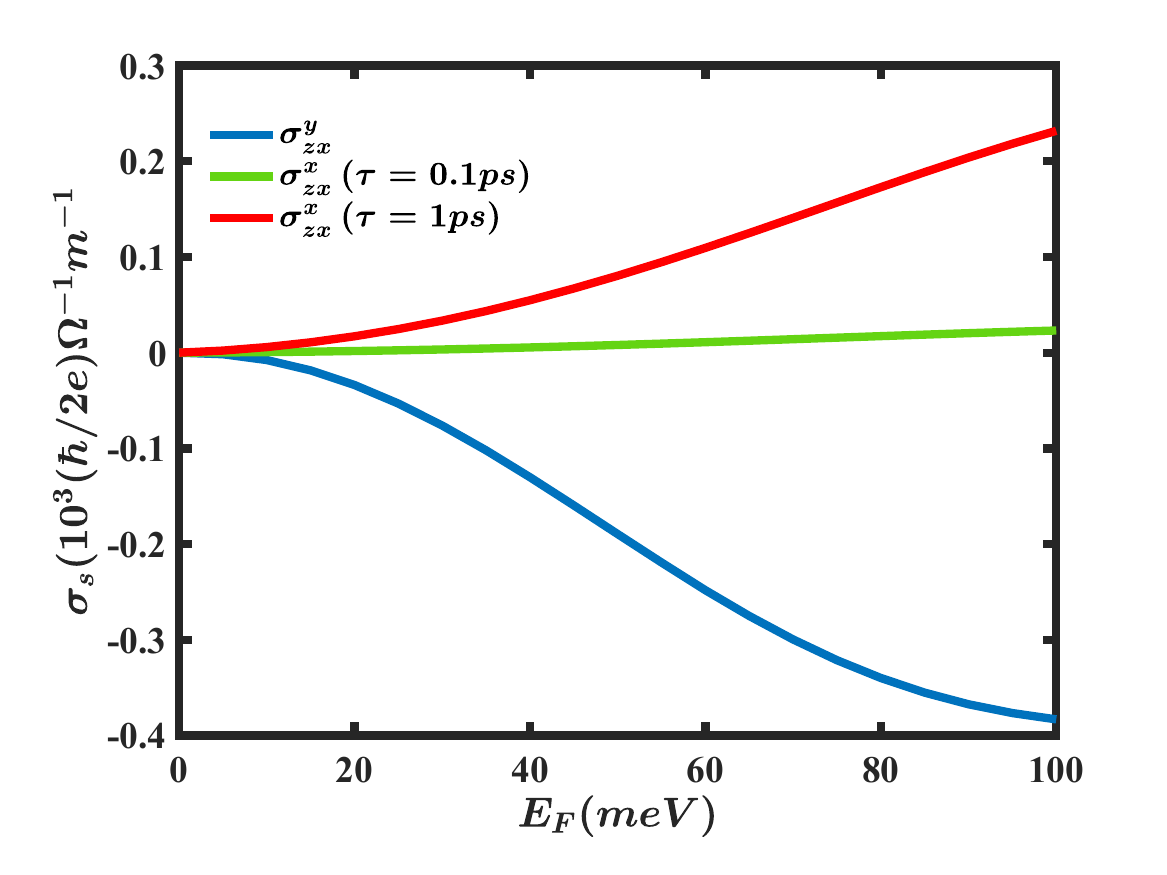}
	\caption{\label{Fig2} The total extrinsic spin Hall conductivity $\sigma_s$ vs the Fermi energy $E_F$ for the TI bulk states in Bi$_2$Se$_3$ with Zeeman field $\boldsymbol{m}\parallel \hat{z}$ and $|m|=10$ $\mu$eV (Bi$_2$Se$_3$ parameters from Ref.~\onlinecite{Liu2010}).}
\end{figure}

The extrinsic term from the torque dipole contribution is
\begin{equation}\label{SCcorr}
    I^i_{j}={\rm Tr}\,t_{i, {\bm q}} \bigg(\pd{S_{{\bm q} {\bm Q}}}{Q_j}\bigg)_{{\bm Q} \rightarrow 0}\,,
\end{equation}
where $t_i=\frac{i}{\hbar}[H_0,si]$ and the part of $S_{{\bm q} {\bm Q}}$ that contribute to the spin current is
\begin{equation}
    S_{\bm q Q}^{mn} = -i\hbar\frac{J_{mn}(n_{{\bm q}{\bm Q}})}{(\varepsilon_m - \varepsilon_n)}\,.
\end{equation}
The terms in (\ref{SCcon}) and (\ref{SCcorr}) together give the extrinsic proper spin current. Note that the kinetic equations used in this derivation assume that the system is in the weak scattering limit.\cite{Culcer2017} Hence, this formula can be used generally to calculate extrinsic spin currents in the weak scattering limit for any system that can be described by a single particle Hamiltonian, in this work we focus on spin currents in topological insulators. 

\section{Results}

We solved equations (\ref{KE1}), (\ref{KE2}) and (\ref{QKE}) numerically to find the linear response of the TI bulk states to an electric field and calculated the induced spin currents flowing out-of-plane $\parallel\hat{z}$. We calculated the scattering term $\hat{J}_0$ by first integrating the scattering out, then the scattering in was calculated iteratively until convergence was reached. Further details on this part of the calculation can be found in the supplement.

We calculated the spin conductivities $\sigma^i_{zx}$ where $i=x,y,z$, this means we have a spin current flowing in $\hat{z}$ of spins aligned along $\hat{i}$ in response to an electric field along $\hat{x}$. These are the spin conductivities relevant to spin torques, as we are concerned with spins flowing from the TI into the interface with the magnetic material. For the following discussion we set the Zeeman field $\boldsymbol{m}\parallel\hat{z}$, we discuss results for other Zeeman field orientations later. We found $\mathcal{J}^z_{zx}$ to be exactly 0. However, we found spin currents $\mathcal{J}^x_{zx}$ and $\mathcal{J}^y_{zx}$ to be non zero. 

\subsection{Extrinsic Contributions to the TI SHE}

The primary contributions to the spin current $\mathcal{J}^y_{zx}$ originate from the electric field correction to the scattering as well as the the band structure skew scattering. Interestingly, it has recently been shown that these same mechanisms are also important for the surface state torque.\cite{farokhnezhad2022,farokhnezhad2023} We find that there is also a contribution from the extrinsic off-diagonal elements of the density matrix $S_E$, however this contribution is of a negligible magnitude. This means that the primary contribution to this spin Hall current come from $n_E^0$, hence the size of this spin current should be independent of the impurity density.

\begin{figure}[t!]
	\centering
	\includegraphics[width = \columnwidth]{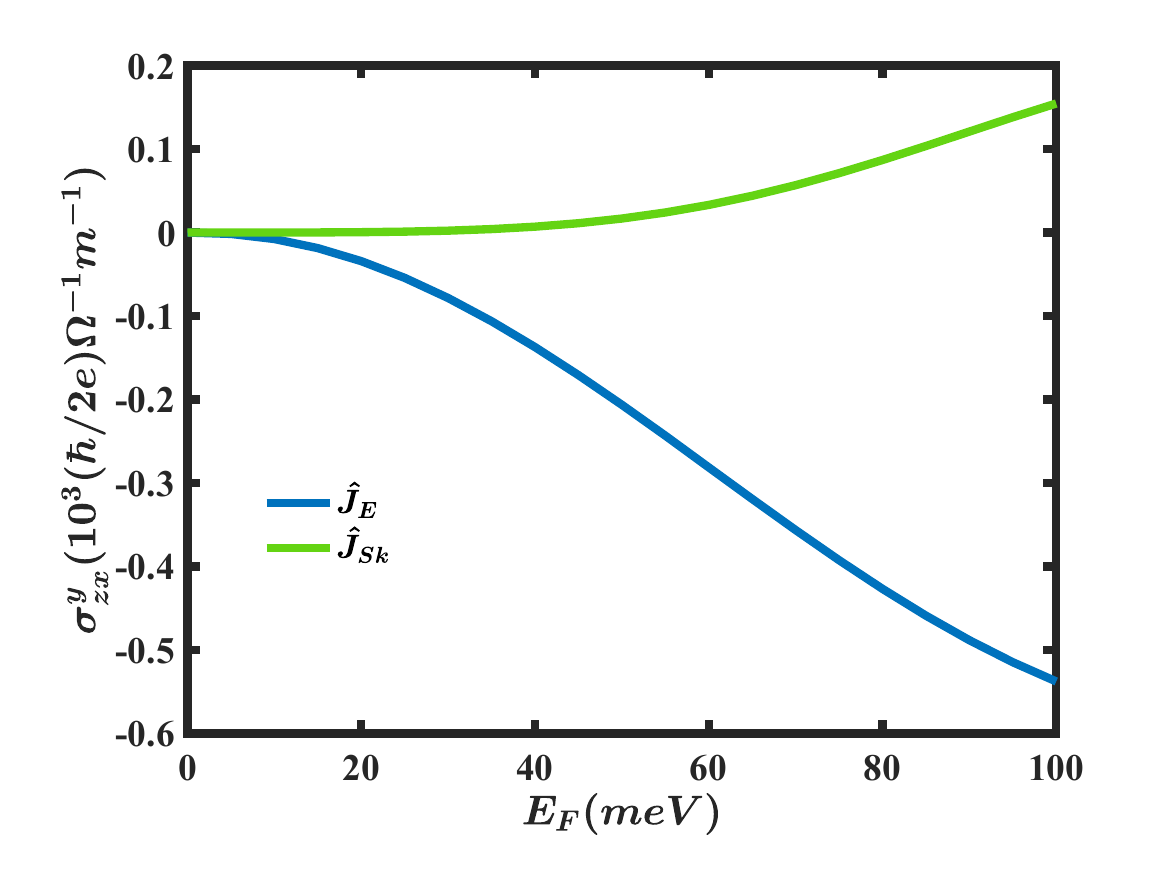}
	\caption{\label{Fig3} The extrinsic spin Hall conductivity $\sigma^{y}_{zx}$ vs the Fermi energy $E_F$ for the TI bulk states due to skew scattering and electric field scattering in Bi$_2$Se$_3$ with Zeeman field $\boldsymbol{m}\parallel \hat{z}$ and $|m|=10$ $\mu$eV (Bi$_2$Se$_3$ parameters from Ref.~\onlinecite{Liu2010}).}
\end{figure}

\begin{figure}[t!]
	\centering
	\includegraphics[width = \columnwidth]{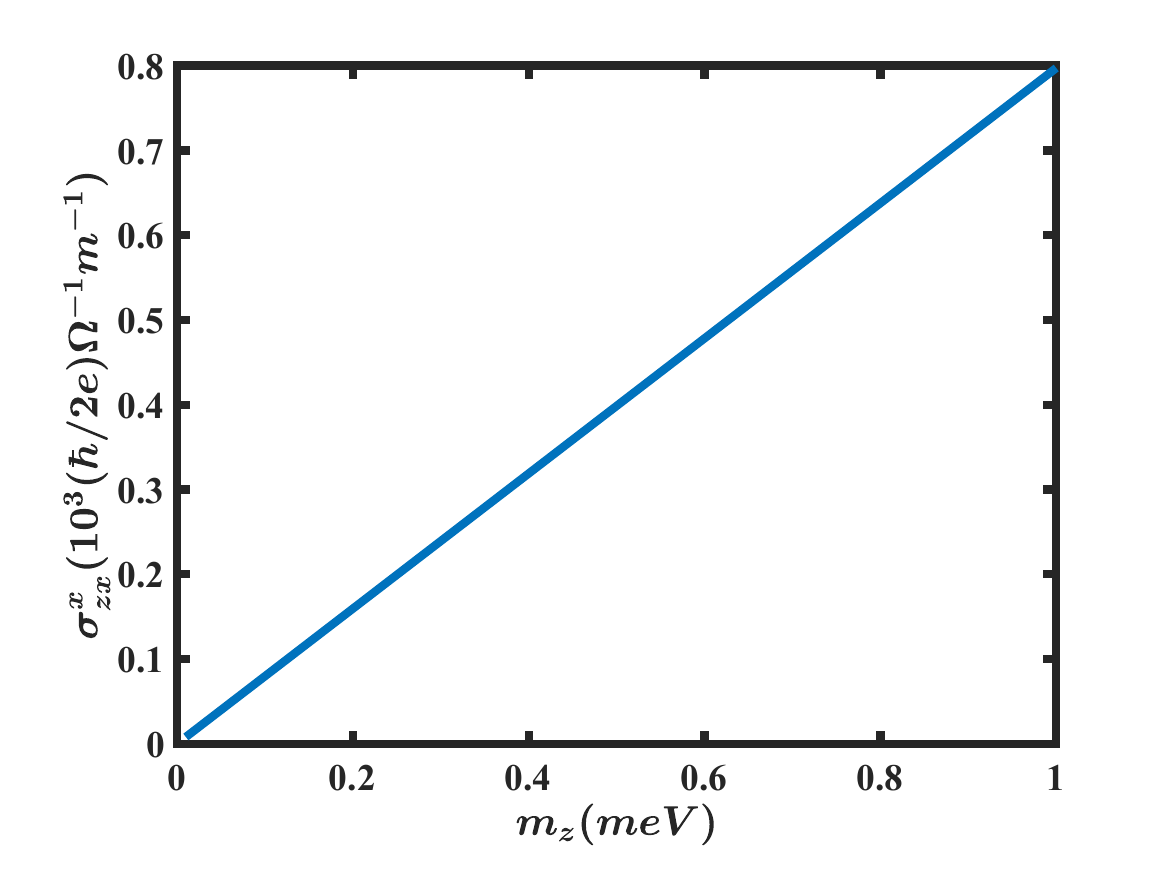}
	\caption{\label{Fig4} The extrinsic spin Hall conductivity $\sigma^x_{zx}$ vs the Zeeman energy $m_z$ for the TI bulk states. For Bi$_2$Se$_3$ with Fermi energy $50$ meV, scattering time $\tau=0.1$ ps and Zeeman field $\boldsymbol{m}\parallel \hat{z}$ (Bi$_2$Se$_3$ parameters from Ref.~\onlinecite{Liu2010}).}
\end{figure}

The spin current $\mathcal{J}^x_{zx}$ is due to the band diagonal part of the density matrix $n_E^{(-1)}$ and can be of a similar order of magnitude to the intrinsic spin current. This spin current does not have any contributions from the electric field scattering, skew scattering, band off-diagonal elements or the torque dipole correction. This means that this spin current is linear in the scattering time and can be enhanced in clean samples with a lower impurity density. Estimations based on experimental results of the bulk conductivity\cite{Jash2021} indicate that Bi$_2$Se$_3$ has a scattering time of order $0.1$ ps. So we chose numbers for the impurity density and scattering potential such that we have a scattering time of the same order of magnitude. Calculations with larger scattering times have also been included to demonstrate the dependence of the spin current on the impurity density. Furthermore, the spin current $\mathcal{J}^x_{zx}$ is linear in the Zeeman energy as shown in Fig. \ref{Fig4}. and for a Zeeman energy of $m_z=1$ meV and scattering time $\tau=0.1$ ps its magnitude is twice the value of the spin current $\mathcal{J}^y_{zx}$. This shows that it is also possible to enhance the extrinsic spin Hall effect in magnetized TIs.

We find that in the TI bulk the spin current due to the torque dipole correction (\ref{SCcorr}) is zero for both spin currents $\mathcal{J}^x_{zx}$ and $\mathcal{J}^y_{zx}$. The spin current from the conventional spin current (\ref{SCcon}) is the only contribution to the extrinsic proper spin current. However, even in this case it is crucial to use the proper spin current and not the conventional spin current as most of the contributions from the torque dipole correction exactly cancel terms in the conventional spin current.\cite{liu2023} Further details on these cancellations can be found in the supplement.

We note that the model Hamiltonian is accurate up to a Fermi energy of around 20 meV and remains reasonably accurate up until 40 meV.\cite{Liu2010} Our results beyond this point should be regarded as approximate: they are included here since experimentally the chemical potential of Bi$_3$Se$_2$ will often be on the order of 100 meV.\cite{bianchi2010,bahramy2012,xia2009}

In FIG. \ref{Fig2}. we plot the total extrinsic spin conductivity vs the Fermi energy. We set the zero of energy at the conduction band minimum. The plot shows that the magnitude of the extrinsic spin conductivities tends to increase monotonically with the Fermi energy. It also shows that the extrinsic spin conductivities are of a comparable magnitude as the intrinsic spin conductivity calculated in Ref. \onlinecite{liu2023}. The spin conductivity $\sigma^x_{zx}$ is plotted for two different scattering times $\tau=0.1,\,1$ ps. Note, the scattering time is dependent on the Fermi energy, these $\tau$ numbers represent the scattering time at $E_F=50$ meV. Since we expect the scattering time to be on the order of magnitude $\tau\sim0.1$ ps, we expect the spin current $\mathcal{J}^x_{zx}$ to be of a negligible order of magnitude compared to $\mathcal{J}^y_{zx}$. However, we do comment that in very clean samples this may not be the case.

FIG. \ref{Fig3}. shows the spin conductivity $\sigma^y_{zx}$ due to the first term in (\ref{SCcon}) from the electric field and skew scattering terms. In this figure we can see that the electric field correction gives the largest contribution to the extrinsic spin current, with the spin current due to skew scattering being roughly an order of magnitude smaller than it at lower Fermi energies, and growing to be about one third the size of it at 100 meV. Each contribution has opposite sign and hence they combine destructively when calculating the total spin current.

\subsection{Analysis of Extrinsic TI Spin Currents}

\begin{table}[t!]
	\centering
	\begin{tabular}{ m{0.1\columnwidth} m{0.2\columnwidth} m{0.2\columnwidth} m{0.1\columnwidth}}
		\hline
		& $m\parallel\hat{x}$ & $m\parallel\hat{y}$ & $m\parallel\hat{z}$\\
		\hline
		$\sigma^x_{zx}$ & $\times$ & $\times$ & \checkmark \\
		\hline
		$\sigma^y_{zx}$ & \checkmark & \checkmark & \checkmark \\
		\hline
		$\sigma^z_{zx}$ & \checkmark & \checkmark & $\times$ \\
		\hline
	\end{tabular}
	\caption{\label{tab1}The zero and non zero spin currents in the TI bulk for different Zeeman field orientations.}
\end{table}

We find the direction of the Zeeman field to significantly effect the magnitude and direction of extrinsic spin currents. This is due to the coupling of the Zeeman terms to spin-orbit terms in our Hamiltonian, as shown in Ref. \onlinecite{cullen2023} the conduction band will have an effective spin-orbit field of $H_c = \frac{\hbar}{2} \, {\bm \sigma} \cdot {\bm \Omega} \equiv \frac{\hbar}{2} \, (\sigma_z \Omega_z + \sigma_+ \Omega_- + \sigma_- \Omega_+)$, where $\sigma_\pm = (\sigma_x \pm i \sigma_y)/2$, and $\Omega_z = - (\mathcal{A}^2k_{\parallel}^2/\hbar\mathcal{M}^2) \, m_z + (\mathcal{A}\mathcal{B}k_z/\hbar\mathcal{M}^2) \, {\bm k}_{\parallel}\cdot{\bm m}_{\parallel}$, and $\Omega_\pm = (\mathcal{A}\mathcal{B} k_z k_{\pm}/\hbar\mathcal{M}^2) \, m_z - (\mathcal{B}^2k_z^2/\hbar\mathcal{M}^2)\, m_\pm + (\mathcal{A}^2k_\mp/\hbar\mathcal{M}^2) \, ({\bm k}\times{\bm m})_z$. We can see how this directly relates to our spin currents linear in the scattering time, for a Zeeman field $\parallel\hat{z}$ we have a spin current $\mathcal{J}^x_{zx}$ which can be directly related to the term $(\mathcal{A}\mathcal{B} k_z k_x/\hbar\mathcal{M}^2) \, m_z \sigma_x$ in the spin-orbit field. Furthermore, for a rotated Zeeman field aligned $\parallel\hat{x}$ we find the spin current linear in the Zeeman energy to have the spin rotated such that we get a spin current of identical magnitude $\mathcal{J}^z_{zx}$ which can be related to the term $(\mathcal{A}\mathcal{B} k_z k_x/\hbar\mathcal{M}^2) \, m_x \sigma_z$. For a Zeeman field $\parallel\hat{y}$ there will be no spin current that is linear in the scattering time, this is consistent with the above analysis as there are no spin orbit terms with $m_y$, $k_x$ and $k_z$. 

For the spin currents independent of the scattering time, we find that although they are largely independent of the magnitude Zeeman field they are dependent on the direction of the Zeeman field. The scattering time independent spin currents can again be described using the effective spin-orbit Hamiltonian. However, it requires a more detailed analysis than was used in the previous paragraph, with reference to the Berry connection and the derivative of the scattering matrix elements, as these are the quantities that appear in the electric field correction to the scattering term $\hat{J}_E$. In the following analysis we consider the electric field to be $\parallel \hat{x}$ and the scattering time independent component of the band diagonal non equillibrium density matrix $n_E \propto \mathcal{R}_x,\, \nabla_{k_x} U_{\boldsymbol{k}^{\prime} \boldsymbol{k}}$. For a Zeeman field $\parallel \hat{x}$, the band diagonal component of the Berry connection $\mathcal{R}_x$ contains a term $(\mathcal{A}^3 \mathcal{B} k_xk_yk_z/4\mathcal{M}^4)\sigma_z$ and, the band diagonal component of the spin operator is $s_{y,d}=(\hbar \mathcal{A}^2 k_x k_y/4 \mathcal{M}^2)\sigma_z$. The band diagonal velocity operator $v_z$ has a factor $\propto k_z \mathbb{I}$. Hence, it is clear that the trace ${\rm Tr}[s_y n_E v_z]$ will be non zero and that there is spin current $\mathcal{J}^y_{zx,m_x}$ that is third order in the spin-orbit field $\propto |\boldsymbol{\Omega}|^3$. When the Zeeman field is aligned $\parallel \hat{y}$, the band diagonal component of the Berry connection $\mathcal{R}_x$ contains a correction of the form $-(\mathcal{A} \mathcal{B} k_z/2\mathcal{M}^2)\sigma_z$. This correction to the Berry connection is due to the Schrieffer-Wolf transform that was used to obtain the effective $2\times2$ Hamiltonian. The leading term in the band diagonal component of the spin operator $s_y$ is $(\hbar/2)\sigma_z$. Hence, the product of these terms with the velocity operator will yield a non zero spin current $\mathcal{J}^y_{zx,m_y}$ that is first order in the spin-orbit field $\propto |\boldsymbol{\Omega}|$. When the Zeeman field is aligned $\parallel \hat{z}$, the band diagonal component of the Berry connection $\mathcal{R}_x$ contains the term $(-k_y/k_\parallel^2+\mathcal{A}^2 k_y/2\mathcal{M}^2)\sigma_z$. Interestingly, $\nabla_{k_x} U_{\boldsymbol{k}^{\prime} \boldsymbol{k}}$ will also yield a term $(k_y/k_\parallel^2)\sigma_z$ that will exactly cancel with the first term from the Berry connection. The band diagonal component of the spin operator is $s_{y,d}=(\hbar \mathcal{A} \mathcal{B} k_y k_z/4 \mathcal{M}^2)\sigma_z$. Hence, the product of these terms with the velocity operator will yield a non zero spin current $\mathcal{J}^y_{zx,m_z}$ that is second order in the spin-orbit field $\propto |\boldsymbol{\Omega}|^2$. Thus, for each of the three orientations of the Zeeman field considered we find that the spin current $\mathcal{J}^y_{zx}$ will be of a different order in the spin-orbit field.

\begin{figure}[t!]
	\centering
	\includegraphics[width = \columnwidth]{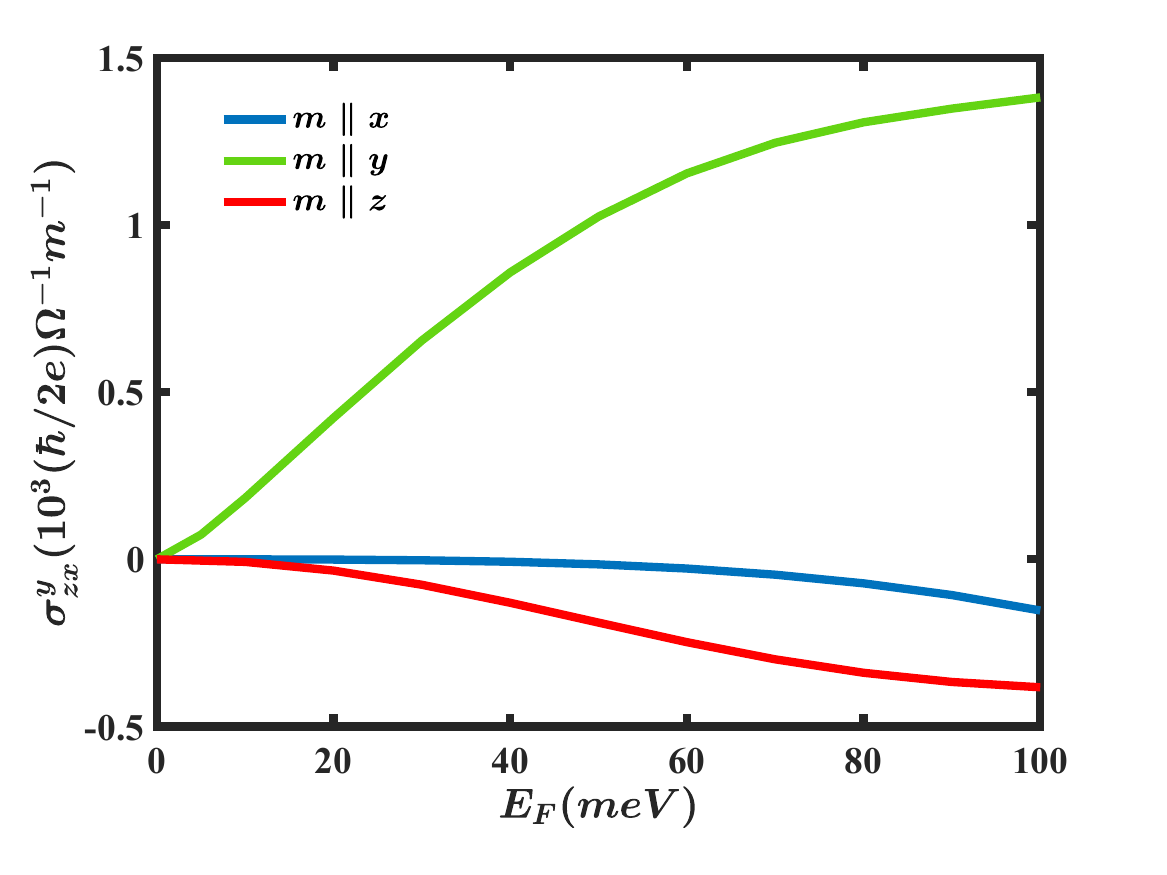}
	\caption{\label{Fig5} The extrinsic spin Hall conductivity $\sigma^y_{zx}$ vs the Fermi energy $E_F$ for the TI bulk states with different Zeeman field orientations. For Bi$_2$Se$_3$ with Zeeman energy $|m|=10$ $\mu$eV (Bi$_2$Se$_3$ parameters from Ref.~\onlinecite{Liu2010}).}
\end{figure}

A summary of the zero and non-zero spin currents can be found in Table \ref{tab1}. As is shown in this table there is an additional non zero spin current $\mathcal{J}^z_{zx}$ for a Zeeman field aligned $\parallel \hat{y}$. This spin current is independent of disorder strength and its size is around three orders of magnitude smaller than the other spin currents we have calculated. We find that the magnitude of this spin current is linear in the Zeeman energy. To describe this spin current we must again refer to the Berry connection and spin operators in the effective spin-orbit Hamiltonian. The berry connection $\mathcal{R}_x$ will contain a component $(\mathcal{A}^2 k_y/2\mathcal{M}^2)\mathbb{I}$. The spin operator $s_z$ will contain a component $(\mathcal{A}\mathcal{B}k_y k_z/2\mathcal{M}^3)m_y\mathbb{I}$, this term is a correction due to the rotation of the spin operator $s_z$ by the Schrieffer-Wolf transform. These terms will yield a spin current $\mathcal{J}^z_{zx}$ that is to second order in the spin-orbit field $\propto |\boldsymbol{\Omega}|^2$. However, this spin current $\mathcal{J}^z_{zx}$ has a factor of $m_y/\mathcal{M}$, which is the ratio of the Zeeman splitting to the band gap, due to this additional factor this spin current will be negligible in most cases.

The model Hamiltonian we used can also describe Bi$_2$Te$_3$ and Sb$_2$Te$_3$. A comparison of the extrinsic spin Hall conductivities of each material can be found in Table \ref{tab2}. These results show that, of these three materials Bi$_2$Se$_3$ should have the largest extrinsic spin Hall effect.

\begin{table}[t!]
	\centering
	\begin{tabular}{ m{0.34\columnwidth} m{0.15\columnwidth} m{0.15\columnwidth} m{0.15\columnwidth}}
		\hline
		& Bi$_2$Se$_3$ & Bi$_2$Te$_3$ & Sb$_2$Te$_3$\\
		\hline
		$\sigma^x_{zx}$ $(\hbar/2e)\Omega^{-1}m^{-1}$ & 8.0 & 1.3 & -0.15 \\
		\hline
		$\sigma^y_{zx}$ $(\hbar/2e)\Omega^{-1}m^{-1}$ & -189.6 & -50.9 & -21.0 \\
		\hline
	\end{tabular}
	\caption{\label{tab2} Extrinsic spin conductivities for Bi$_2$Se$_3$, Bi$_2$Te$_3$ and Sb$_2$Te$_3$, calculated for Fermi energy $E_F=50$ meV, scattering time $\tau=0.1$ ps and Zeeman field $\boldsymbol{m}\parallel\hat{z}$ with $|m|=10$ $\mu$eV (material parameters from Ref.~\onlinecite{Liu2010}).}
\end{table}

\subsection{Hexagonal Warping}

\begin{figure}[t!]
	\centering
	\includegraphics[width = \columnwidth]{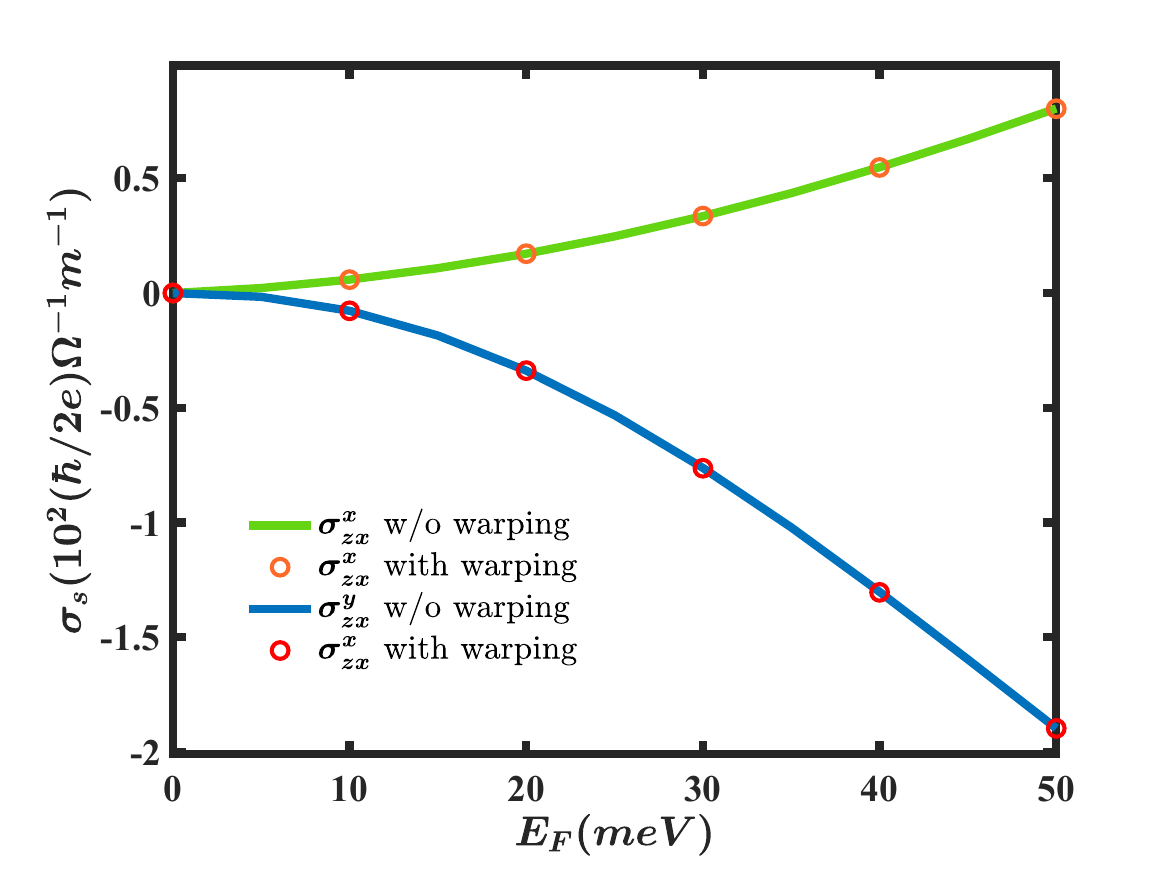}
	\caption{\label{Fig6} The extrinsic spin Hall conductivity $\sigma_s$ vs the Fermi energy $E_F$ for the TI bulk states due to electric field scattering with and without hexagonal warping included. For Bi$_2$Se$_3$ with scattering time $\tau=1$ ps, Zeeman field $\boldsymbol{m}\parallel \hat{z}$ and $|m|=10$ $\mu$eV (Bi$_2$Se$_3$ parameters from Ref.~\onlinecite{Liu2010}).}
\end{figure}

Up to this point our calculations have ignored the hexagonal warping terms that appear in topological insulators. These extra warping terms in the Hamiltonian have the form
\begin{equation}
    H_w=\frac{R_1}{2}(k_+^3+k_-^3)\mathbb{I}\otimes\sigma_y+i\frac{R_2}{2}(k_+^3-k_-^3)\sigma_z\otimes\sigma_x\,,
\end{equation}
where, $R_1$ and $R_2$ are material specific parameters. These terms were ignored because they increase the complexity of the dispersion and eigenstates. However, we did do some calculations with them to approximate their effect on the proper spin current. We plot the extrinsic spin conductivity calculated with and without the warping terms included in FIG. \ref{Fig6}. The plot shows that the warping terms do not effect the magnitude of the spin current at lower Fermi energies where our model is valid.

\section{Discussion}

Here we have demonstrated a straightforward method for calculating the proper spin current due to impurity scattering. This, along with our previous work on intrinsic spin currents,\cite{liu2023} provide straightforward formulae for calculating the total proper spin current. This method can be applied generally to other systems of interest, and can be applied to any system, for example van der Waals heterostructures and other exotic materials.\cite{Giant-SHE-Metal-Adv-Mat-2022, muduli2019evaluation, zhao2020observation, avsar2014spin, benitez2020tunable, safeer2019room, safeer2020spin, DASTGEER2022, herling2020gate, ghiasi2019charge, hoque2022charge} In this work we applied our method for calculating the proper spin current to topological insulators and calculated the spin conductivity of Bi$_2$Se$_3$, Bi$_2$Te$_3$ and Sb$_2$Te$_3$ both with and without hexagonal warping. We find that the "side jump" scattering term $\hat J_E$ is the dominant contribution to the extrinsic spin Hall effect. Furthermore, we find that the extrinsic spin Hall effect is of a similar magnitude to the intrinsic Hall effect in topological insulators,\cite{liu2023} this is demonstrated in FIG.~\ref{Fig1}, where both extrinsic and intrinsic spin conductivities are plotted. We find that when the Fermi energy is in the conduction band the spin conductivity of the bulk TI states is $\sigma^y_{zx}\sim 10^3(\hbar/2e)\Omega^{-1}\text{m}^{-1}$. 

In the present study, we focused solely on short-ranged scalar impurities. However, our methodology allows for the inclusion of more complex scattering potentials and, can be applied to further investigations of the spin Hall effect, incorporating spin-orbit scattering beyond what we have considered. Our spin current equations are applicable to both 3D and 2D systems. In the case of 2D systems, additional consideration may be required for weak localization effects, which can be incorporated through modifications to the linear response formalism outlined here.\cite{liu2023coherent}

\begin{figure}[t!]
	\centering
	\includegraphics[width = \columnwidth]{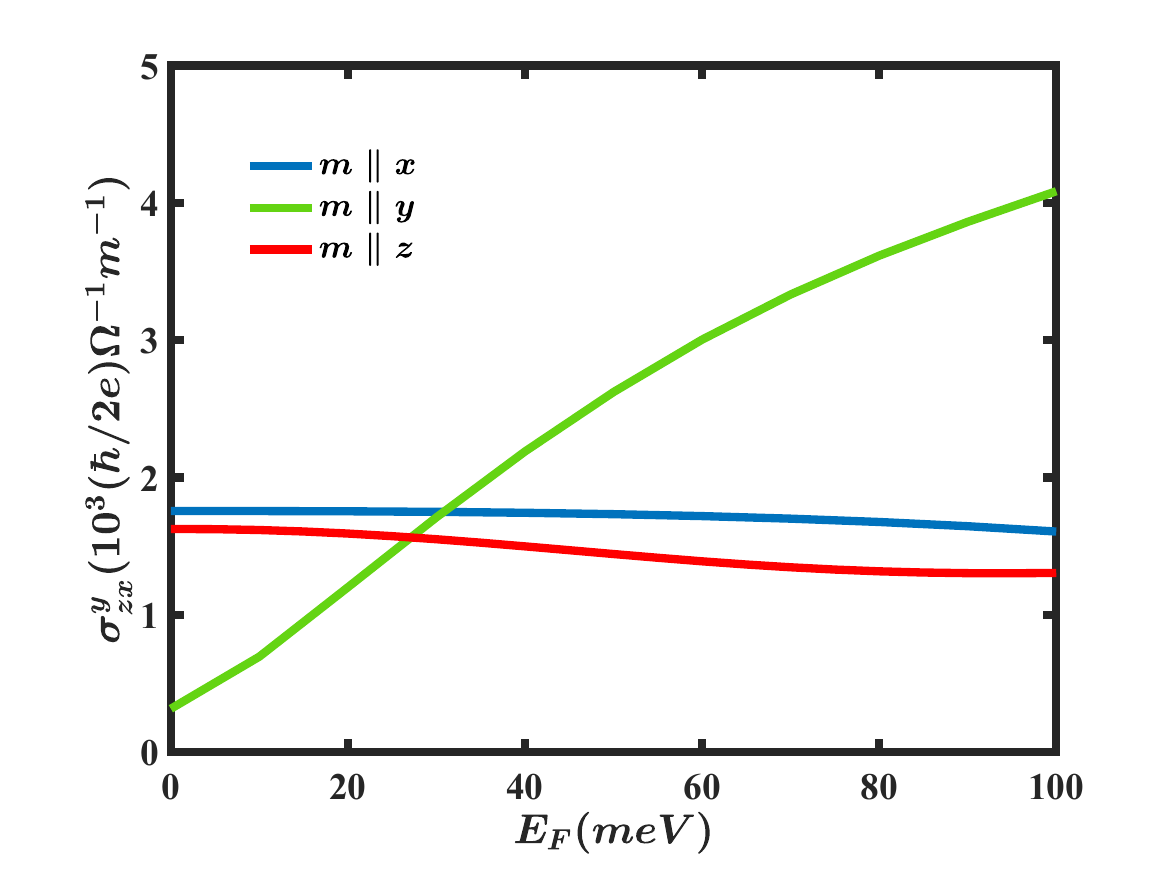}
	\caption{\label{Fig7} The total spin Hall conductivity $\sigma^{y}_{zx}$ (intrinsic and extrinsic) vs the Fermi energy $E_F$ for the TI bulk states in Bi$_2$Se$_3$ with Zeeman energy $|m|=10$ $\mu$eV (Bi$_2$Se$_3$ parameters from Ref.~\onlinecite{Liu2010}).}
\end{figure}

We employed an effective $2\times2$ spin-orbit Hamiltonian used in Ref.~\onlinecite{cullen2023} for our analysis. We find that the spin currents linear in the scattering time, $\mathcal{J}^x_{zx}$ where $\boldsymbol{m}\parallel\hat{z}$ and $\mathcal{J}^z_{zx}$ where $\boldsymbol{m}\parallel\hat{x}$, have identical magnitude, this is the case in both our analysis and in the numerical calculation. Conversely, the spin currents independent of disorder strength are of different orders in the spin-orbit field for different orientations of the Zeeman field, this implies that at low Fermi energies $E_F<10$ meV, there will be large differences in the magnitude of the spin current $\mathcal{J}^y_{zx}$. From our analysis we expect $|\mathcal{J}^y_{zx,m_y}| \gg |\mathcal{J}^y_{zx,m_x}| \gg |\mathcal{J}^y_{zx,m_z}|$ and that $\mathcal{J}^y_{zx,m_y}$ will have opposite sign to $\mathcal{J}^y_{zx,m_x}$ and $\mathcal{J}^y_{zx,m_z}$. This is consistent with our numerical results at low Fermi energies as shown in FIG. \ref{Fig5}, where the extrinsic spin conductivity $\sigma^y_{zx}$ has been plotted for three different Zeeman field directions. Furthermore, we find that even at larger Fermi energies beyond where the effective Hamiltonian is valid, this hierarchy in spin current magnitudes for each Zeeman field orientation still exists. For Zeeman fields oriented $\parallel\hat{z}$ or $\hat{x}$ the intrinsic and extrinsic spin conductivities $\sigma^y_{zx}$ have opposite signs and, the magnitude of the intrinsic spin conductivity increases slower than the extrinsic spin conductivity while increasing the Fermi energy. Hence, for these cases the spin conductivity will be at a maximum when the Fermi energy is in the band gap, though this may not be the case for larger Fermi energies beyond where our model is accurate. Conversely, for a Zeeman field $\parallel \hat{y}$ the extrinsic and intrinsic spin conductivities will have the same sign and add constructively. This is demonstrated in FIG. \ref{Fig7} in which we have plotted the total spin conductivity $\sigma^y_{zx}$ including both intrinsic and extrinsic contributions. 

It should be noted that the dependence of the spin current on the direction of the spin-orbit field is a smoking gun for the measurement of the extrinsic spin Hall effect. This can be tested using a TI/FM sample by varying the orientation of a small external magnetic field and measuring the changes in the size of the spin torque. Although, the orbital effects of a magnetic field are not considered in this work as long as the magnetic field is small enough any orbital effects should be negligible. The dependence of the spin Hall effect on the Zeeman field and the spin-orbit field $\boldsymbol{\Omega}$ has interesting parallels to the previous work on the bulk spin transfer torque.\cite{cullen2023}

The results we have presented for TIs have in-plane $x$ - $y$ symmetry, for example our results with $m\parallel\hat{z}$ are the same for the electric fields oriented $\parallel \hat{x}$ and $\parallel \hat{y}$, we find that $\sigma^x_{zx}=\sigma^y_{zy}$ and $\sigma^y_{zx}=-\sigma^x_{zy}$. Remarkably, this holds even when including the hexagonal warping terms that remove the in-plane symmetry from the Hamiltonian. The spin conductivities will differ for an out-of-plane electric field, we have not studied such a configuration here as it is not often used in spin torque devices.

Spin torques in topological insulator spin torque devices are known to be driven by various mechanisms, including the spin Hall effect in the bulk, \cite{Chang2015, Ghosh2018, Ado2017} the Rashba-Edelstein effect (REE) in the surface states, and the spin transfer torque (STT) due to the proximity effect with the adjacent magnetic layer.\cite{Kurebayashi2019,cullen2023} In the field of TI spin torques, a crucial yet unresolved question pertains to determining the relative magnitude of each spin torque mechanism. The Rashba-Edelstein effect is known to be sizable at the surface topological insulators.\cite{sakai2014, Fischer2016, Ndiaye2017, Chang2015,siu2018} However, it has been shown that the chemical potential lies in the bulk TI conduction band for most TI/FM devices,\cite{Zhang2016, Marmolejo-Tejada2017} and that bulk transport dominates in a certain parameter regime.\cite{Jash2021} So, spin torques due to the bulk states are unable to be neglected. Furthermore, recently the spin density due to STT mechanism in the bulk states was shown to potentially be of the same order of magnitude as the spin density generated by the REE in the surface states.\cite{cullen2023} Given the low conductivity of TI samples it can be inferred that samples are generally quite dirty. Hence, it is also important to consider extrinsic effects in these materials. We find our calculated spin conductivities to be roughly one to two orders of magnitude smaller than those recorded in experiment $\sigma_s=0.15-2 \times 10^5 (\hbar/2e) \Omega^{-1}\text{m}^{-1}$.\cite{Mellnik2014, Han2017, Dc2018, Wang2017} This indicates that the large charge to spin conversion efficiency of TIs measured in experiment is largely due to other spin torque mechanisms and not the spin Hall effect. Although experimental works calculate the spin conductivity, which is related to the spin Hall effect, what is measured is the spin torque which has contributions from spin polarizations generated via other mechanisms.\cite{Wang2018} 

It must be noted that directly relating the spin current to a spin torque is difficult, due to the complexity of accurately calculating the effect of the TI/FM interface on the spin current. Our calculations are of spin currents in the bulk of the TI and, hence do not directly relate to the spin conductivity measured in experiment. Spin memory loss\cite{gupta2020disorder, Dolui2017} can occur at the interface, which would reduce the effect of the bulk spin Hall effect on the spin torque, supporting the view proposed here that the SHE in negligible in topological insulator spin torques.

Our calculations show that the extrinsic spin Hall effect will generate both damping and field-like spin currents. Furthermore, we predict these spin currents can be of a similar magnitude. However, they exhibit specific differences, for example the field-like current is independent of impurity density whereas the damping-like current is dependent on the impurity density. The damping-like current is linear in the scattering time and hence should be larger in cleaner samples. The field-like current has no such dependence, furthermore the intrinsic spin conductivity is also purely field-like. Hence, neglecting interface effects, the field-like contribution to the spin torque due to the spin Hall effect will be entirely independent of the impurity density. Although we predict the damping-like spin current  to likely be small. It has the possibility to be of a comparable size to the field-like current in very clean samples. Hence, if the spin Hall effect is a significant part of the TI spin torque we may expect the spin conductivity to be significantly smaller in more disordered samples. However, a study of spin torques in sputtered topological insulators measured an exceedingly large spin conductivity of $\sigma_s=1.5 \times10^5 (\hbar/2e) \Omega^{-1}\text{m}^{-1}$.\cite{Dc2018} This further indicates that the spin Hall effect is likely negligible in TI spin torques.

The results in this paper combined with the results in Refs. \cite{cullen2023,liu2023} give a comprehensive analysis of spin torques stemming from the TI bulk states. The bulk states give two contributions to the spin torque, spin currents from the spin Hall effect and spin polarizations due to the spin transfer torque mechanism. The spin Hall effect calculated is small and contributes negligibly to the total spin torque. The bulk spin transfer torque calculated with an idealised model is negligible when compared to the surface state torque, but can potentially be large in real samples. So, the only bulk TI spin torque contribution that can compete with the Rashba-Edelstein effect in the surface states is the spin transfer torque. This is consistent with experimental results that find the spin torque efficiency increase for thinner TI samples,\cite{Wang2017} which implies that states at or near the interface are the dominant contribution to the spin torque and that purely bulk contributions are negligible. These results do not preclude the possibility of a substantial spin transfer torque, since spin polarizations from this effect are localized near the interface where there is a proximity-induced magnetization in the topological insulator.

\section{Conclusion}
In conclusion, we formulated a fully quantum mechanical way of calculating the extrinsic spin Hall effect. We applied it to topological insulator systems to investigate the extrinsic spin Hall effects role in topological insulator spin torques. We found the extrinsic spin conductivity to be roughly 2 orders of magnitude smaller than those reported in experiment and concluded that its role in spin torques is negligible.

\section*{Acknowledgements}
This project was supported by the Australian Research Council Future Fellowship FT190100062.\\
JHC acknowledges support from an Australian Government Research Training Program (RTP) Scholarship.

\newpage

\setcounter{section}{0}

\onecolumngrid

\section{Evaluating the band-diagonal scattering term}
Here we will outline how the band-diagonal scattering term was evaluated. The the band-diagonal part of the kinetic equation in the steady state is
\begin{equation}
    \left[\hat J_0 (n_E)\right]_{nn} = D_{nn}\,.
\end{equation} 
Where $D_{nn}$ is the driving term. The scattering term in the Born approximation is
\begin{equation}
    J(n_{E})^{mm^\prime} = \frac{\pi n_i}{\hbar} \sum_{n,k^\prime}  \mathcal{U}^{mn}_{k k^\prime}\mathcal{U}^{nm^\prime}_{k^\prime k}\left[(n^m_{\boldsymbol{k}} - n^n_{\boldsymbol{k}^\prime}) \delta(\epsilon^n_{k^\prime}-\epsilon^m_{k})+(n^{m^\prime}_{\boldsymbol{k}} - n^n_{\boldsymbol{k}^\prime}) \delta(\epsilon^n_{k^\prime}-\epsilon^{m^\prime}_{k})\right]
\end{equation}
In order to calculate this we split it up into two parts scattering in and scattering out. The scattering in term contains the parts with $n_{\boldsymbol{k}^\prime}$,
\begin{equation}
    J(n_{E})^{mm^\prime}_{in} = -\frac{\pi n_i}{\hbar} \sum_{n,k^\prime}  \mathcal{U}^{mn}_{k k^\prime}\mathcal{U}^{nm^\prime}_{k^\prime k}\left[n^n_{\boldsymbol{k}^\prime} \delta(\epsilon^n_{k^\prime}-\epsilon^m_{k})+n^n_{\boldsymbol{k}^\prime} \delta(\epsilon^n_{k^\prime}-\epsilon^{m^\prime}_{k})\right]\,,
\end{equation}
and the scattering out term contains the parts with $n_{\boldsymbol{k}}$,
\begin{equation}
    J(n_{E})^{mm^\prime}_{out} = \frac{\pi n_i}{\hbar} \sum_{n,k^\prime}  \mathcal{U}^{mn}_{k k^\prime}\mathcal{U}^{nm^\prime}_{k^\prime k}\left[n^m_{\boldsymbol{k}} \delta(\epsilon^n_{k^\prime}-\epsilon^m_{k})+n^{m^\prime}_{\boldsymbol{k}} \delta(\epsilon^n_{k^\prime}-\epsilon^{m^\prime}_{k})\right]\,.
\end{equation}
Now the kinetic equation can be rewritten as
\begin{equation}
    \hat J_{0,in} (n_E) + \hat J_{0,out} (n_E) = D_{nn}\,,
\end{equation} 
The scattering out term is straightforward to integrate. Additionally, the scattering in term is usually much smaller than the scattering out, so we can use this to evaluate the kinetic equation iteratively. First we rearrange the kinetic equation
\begin{equation}
    \frac{n_E}{\tau(\boldsymbol{k})} = D_{nn}-\hat J_{0,in} (n_E) \,.
\end{equation} 
Where $\tau(\boldsymbol{k})$ is the scattering time, found by simply integrating the scattering out term. Now, we can solve this iteratively by first calculating
\begin{equation}
    n_E^0 = \tau(\boldsymbol{k})D_{nn}\,,
\end{equation} 
then iterating using the following equation
\begin{equation}
    n_E^n = -\tau(\boldsymbol{k})\hat J_{0,in} (n_E^{(n-1)}) \,.
\end{equation} 
until convergence is reached.

\section{Born approximation scattering term to first order in $Q$}
Here we expand the scattering term to linear order in $\boldsymbol{Q}$ and, show that it simplifies to the regular scattering term. The scattering term in the $\boldsymbol{Q}$ expansion is
\begin{equation}
    J(\langle \rho\rangle)^{mm^\prime}_{q_+q_-} = \frac{1}{\hbar^2}\int_0^\infty dt^\prime \langle[U,[e^{-iH_0t^\prime/\hbar}Ue^{iH_0t^\prime/\hbar},\langle\rho\rangle]]\rangle^{mm^\prime}_{q_+q_-}
\end{equation}
First we will consider $\langle \rho \rangle = n_{\boldsymbol{q}\boldsymbol{Q}}$ and only the band diagonal part of the scattering term. Expanding the commutators this makes the scattering term
\begin{equation}
\begin{aligned}
    J(\langle \rho\rangle)^{mm}_{q_+q_-} =& \frac{1}{\hbar^2}\int_0^\infty dt^\prime \sum_{n,k^\prime} \langle U^{mn}_{q_+ k^\prime}U^{nm}_{k^\prime q_+}\rangle \left(n^m_{\boldsymbol{q}\boldsymbol{Q}} e^{-i\epsilon^n_{k^\prime} t^\prime/\hbar}e^{i\epsilon^m_{q_+}t^\prime/\hbar} + n^m_{\boldsymbol{q}\boldsymbol{Q}} e^{-i\epsilon^m_{q_-}t^\prime/\hbar}e^{i\epsilon^n_{k^\prime} t^\prime/\hbar}\right)\\ 
    &-\sum_{n,q^\prime,Q^\prime}\langle U^{mn}_{q_+ q_+^\prime}U^{nm}_{q_-^\prime q_-}\rangle \left(n^n_{\boldsymbol{q}^\prime \boldsymbol{Q}^\prime} e^{-i\epsilon^n_{q_-^\prime} t^\prime/\hbar}e^{i\epsilon^m_{q_-}t^\prime/\hbar} + n^n_{\boldsymbol{q}^\prime \boldsymbol{Q}^\prime} e^{-i\epsilon^n_{q_+^\prime} t^\prime/\hbar}e^{i\epsilon^m_{q_+}t^\prime/\hbar}\right)
\end{aligned}
\end{equation}
Carrying out the time integral and the disorder average we get
\begin{equation}
\begin{aligned}
    J(\langle \rho\rangle)^{mm}_{q_+q_-} =& \frac{\pi n_i}{\hbar} \sum_{n,k^\prime}  \mathcal{U}^{mn}_{q_+ k^\prime}\mathcal{U}^{nm}_{k^\prime q_+} \left(n^m_{\boldsymbol{q}\boldsymbol{Q}} \delta(\epsilon^n_{k^\prime}-\epsilon^m_{q_+}) + n^m_{\boldsymbol{q}\boldsymbol{Q}} \delta(\epsilon^n_{k^\prime}-\epsilon^m_{q_-})\right)\\ 
    &-\sum_{n,q^\prime,Q^\prime}\mathcal{U}^{mn}_{q_+ q_+^\prime}\mathcal{U}^{nm}_{q_-^\prime q_-}\delta(\boldsymbol{Q}-\boldsymbol{Q}^\prime) \left(n^n_{\boldsymbol{q}^\prime \boldsymbol{Q}^\prime} \delta(\epsilon^n_{q_-^\prime}-\epsilon^m_{q_-}) + n^n_{\boldsymbol{q}^\prime \boldsymbol{Q}^\prime} \delta(\epsilon^n_{q_+^\prime} -\epsilon^m_{q_+})\right)
\end{aligned}
\end{equation}
The extra delta function appears due to the disorder average. To linear order in $\boldsymbol{Q}$ this simplifies to the normal scattering term
\begin{equation}
    J(\langle \rho\rangle)^{mm}_{q_+q_-} = \frac{2 \pi n_i}{\hbar} \sum_{n,q^\prime}  \mathcal{U}^{mn}_{q k^\prime}\mathcal{U}^{nm}_{q^\prime q}(n^m_{\boldsymbol{q}\boldsymbol{Q}} - n^n_{\boldsymbol{q}^\prime\boldsymbol{Q}}) \delta(\epsilon^n_{q^\prime}-\epsilon^m_{q})
\end{equation}
Now we must check the off-diagonal scattering term
\begin{equation}
\begin{aligned}
    J(n_{\boldsymbol{q}\boldsymbol{Q}})^{mm^\prime}_{q_+q_-} =& \frac{\pi n_i}{\hbar} \sum_{n,k^\prime}  \mathcal{U}^{mn}_{q_+ k^\prime}\mathcal{U}^{nm^\prime}_{k^\prime q_+} n^{m^\prime}_{\boldsymbol{q}\boldsymbol{Q}} \delta(\epsilon^n_{k^\prime}-\epsilon^{m^\prime}_{q_+}) + \mathcal{U}^{mn}_{q_- k^\prime}\mathcal{U}^{nm^\prime}_{k^\prime q_-} n^m_{\boldsymbol{q}\boldsymbol{Q}} \delta(\epsilon^n_{k^\prime}-\epsilon^m_{q_-})\\ 
    &-\sum_{n,q^\prime,Q^\prime}\mathcal{U}^{mn}_{q_+ q_+^\prime}\mathcal{U}^{nm^\prime}_{q_-^\prime q_-}\delta(\boldsymbol{Q}-\boldsymbol{Q}^\prime) \left(n^n_{\boldsymbol{q}^\prime \boldsymbol{Q}^\prime} \delta(\epsilon^n_{q_-^\prime}-\epsilon^{m^\prime}_{q_-}) + n^n_{\boldsymbol{q}^\prime \boldsymbol{Q}^\prime} \delta(\epsilon^n_{q_+^\prime} -\epsilon^m_{q_+})\right)
\end{aligned}
\end{equation}
Again to linear order in $\boldsymbol{Q}$ this simplifies to the normal scattering term
\begin{equation}
    J(n_{\boldsymbol{q}\boldsymbol{Q}})^{mm^\prime}_{q_+q_-} = \frac{\pi n_i}{\hbar} \sum_{n,q^\prime}  \mathcal{U}^{mn}_{q q^\prime}\mathcal{U}^{nm^\prime}_{q^\prime q}\left[(n^m_{\boldsymbol{q}\boldsymbol{Q}} - n^n_{\boldsymbol{q}^\prime\boldsymbol{Q}}) \delta(\epsilon^n_{q^\prime}-\epsilon^m_{q})+(n^{m^\prime}_{\boldsymbol{q}\boldsymbol{Q}} - n^n_{\boldsymbol{q}^\prime\boldsymbol{Q}}) \delta(\epsilon^n_{q^\prime}-\epsilon^{m^\prime}_{q})\right]
\end{equation}

\section{Electric field and skew scattering terms}
Here we define the spin-orbit scattering terms we calculated. The first is the electric field correction to the scattering term, which is considered to be a side jump mechanism, the second is the band structure skew scattering term. The electric field correction to the scattering term is
\begin{equation}
    \begin{aligned}
    {\left[J_E\left(n_{F D}\right)\right]_k^m } & =\frac{2 \pi}{\hbar} \frac{\partial n_{F D}\left(\epsilon_k^m\right)}{\partial \epsilon_k^m} e E \cdot \sum_{\boldsymbol{k}^{\prime}}\left\langle U_{\boldsymbol{k} \boldsymbol{k}^{\prime}}^{m m} U_{\boldsymbol{k}^{\prime} \boldsymbol{k}}^{m m}\right\rangle\left[\mathcal{R}_{\boldsymbol{k}^{\prime}}^{m m}-\mathcal{R}_{\boldsymbol{k}}^{m m}\right] \delta\left(\epsilon_{\boldsymbol{k}^{\prime}}^m-\epsilon_{\boldsymbol{k}}^m\right) \\
    & +\frac{2 \pi}{\hbar} \frac{\partial n_{F D}\left(\epsilon_{\boldsymbol{k}}^m\right)}{\partial \epsilon_{\boldsymbol{k}}^m} e \boldsymbol{E} \cdot \sum_{m^{\prime} \boldsymbol{k}^{\prime}} \operatorname{Im}\left\{\left\langle\left[\left(\nabla_{\boldsymbol{k}}+\nabla_{\boldsymbol{k}^{\prime}}\right) U_{\boldsymbol{k} \boldsymbol{k}^{\prime}}^{m m^{\prime}}\right] U_{\boldsymbol{k}^{\prime} \boldsymbol{k}}^{m^{\prime} m}\right\rangle\right\} \delta\left(\epsilon_{\boldsymbol{k}^{\prime}}^{m^{\prime}}-\epsilon_{\boldsymbol{k}}^m\right).
    \end{aligned}
\end{equation}
Where $n_{FD}$ is the equillibrium density matrix. The skew scattering term is
\begin{equation}
    \begin{aligned}
     \left[J_{s k}\left(n_E^{(-1)}\right)\right]_k^m=&\frac{2 \pi^2}{\hbar} \sum_{m^{\prime} m^{\prime \prime} n k^{\prime} \boldsymbol{k}^{\prime \prime}} \operatorname{Im}\left[\frac{\left\langle U_{k k^{\prime}}^{m m^{\prime \prime}} U_{k^{\prime} \boldsymbol{k}}^{m^{\prime} m}\right\rangle\left\langle U_{k^{\prime} k^{\prime \prime}}^{m^{\prime \prime} n} U_{\boldsymbol{k}^{\prime \prime} \boldsymbol{k}^{\prime}}^{n m^{\prime}}\right\rangle}{\left(\epsilon_{\boldsymbol{k}^{\prime}}^{m^{\prime \prime}}-\epsilon_{\boldsymbol{k}^{\prime}}^{m^{\prime}}\right)}\right] \\
    & \left\{\left(n_{E \boldsymbol{k}^{\prime}}^{m^{\prime}(-1)}-n_{E \boldsymbol{k}^{\prime \prime}}^{n(-1)}\right) \delta(\epsilon_{\boldsymbol{k}^{\prime}}^{m^{\prime}}-\epsilon_{\boldsymbol{k}^{\prime \prime}}^n)+\left(n_{E \boldsymbol{k}^{\prime}}^{m^{\prime \prime}(-1)}-n_{E \boldsymbol{k}^{\prime \prime}}^{n(-1)}\right) \delta(\epsilon_{\boldsymbol{k}^{\prime \prime}}^n-\epsilon_{\boldsymbol{k}^{\prime}}^{m^{\prime \prime}})\right\} \delta(\epsilon_{\boldsymbol{k}^{\prime}}^{m^{\prime \prime}}-\epsilon_{\boldsymbol{k}}^m) \\
    & -\frac{2 \pi^2}{\hbar} \sum_{m^{\prime} m^{\prime \prime} n k^{\prime} k^{\prime \prime}} \operatorname{Im}\left[\frac{\left\langle U_{\boldsymbol{k} \boldsymbol{k}^{\prime}}^{m^{\prime \prime} m^{\prime}} U_{\boldsymbol{k}^{\prime} \boldsymbol{k}}^{m^{\prime} m}\right\rangle\left\langle U_{\boldsymbol{k} \boldsymbol{k}^{\prime \prime}}^{m n} U_{\boldsymbol{k}^{\prime \prime} \boldsymbol{k}}^{n m^{\prime \prime}}\right\rangle}{\left(\epsilon_k^m-\epsilon_{\boldsymbol{k}}^{m^{\prime \prime}}\right)}\right] \\
    & \left\{\left(n_{E \boldsymbol{k}}^{m^{\prime \prime}(-1)}-n_{E \boldsymbol{k}^{\prime \prime}}^{n(-1)}\right) \delta\left(\epsilon_{\boldsymbol{k}}^{m^{\prime \prime}}-\epsilon_{\boldsymbol{k}^{\prime \prime}}^n)+\left(n_{\boldsymbol{E} \boldsymbol{k}}^{m(-1)}-n_{E \boldsymbol{k}^{\prime \prime}}^{n(-1)}\right) \delta(\epsilon_{\boldsymbol{k}^{\prime \prime}}^n-\epsilon_{\boldsymbol{k}}^m\right)\right\} \delta(\epsilon_{\boldsymbol{k}^{\prime}}^{m^{\prime}}-\epsilon_{\boldsymbol{k}}^{m^{\prime \prime}}).
    \end{aligned}
\end{equation}

\section{Explicit evaluation of the conserved spin current (off-diagonal terms)}
For this we use the energy eigenstate basis and use the notation in Ref. \onlinecite{Culcer2017}. We will start with the conventional spin current. First note that
\begin{equation}
		\displaystyle J^i_j = \frac{1}{2} \, {\rm Tr} \, \rho \, \{ s_i, v_j \} = \frac{1}{2} \, {\rm Tr} \, s_i \, \{ v_j, \rho \}. 
\end{equation}
Split everything into band-diagonal $d$ and band-off-diagonal $od$ contributions, and recall that the band-diagonal part of the density matrix is called $n$ and the band off-diagonal part is called $S$. First consider the contribution coming from the band-diagonal matrix elements of the spin operator $s_i^d$. Because the trace selects the diagonal elements of the total matrix, $s_i^d$ goes with $\{ v_j, \rho \}^d$, as follows:
\begin{equation}
	\displaystyle J^i_{j, 1} = \frac{1}{2} \, {\rm Tr} \, s_i^d \, \{ v_j, \rho \}^d = {\rm Tr} \, s_i^d \,  v_j^d n_E + \frac{1}{2} \, {\rm Tr} \, s_i^d \, \{ v_j^{od}, S_E \}.
\end{equation}
The term $\{ v_j^{od}, S_E \}_d$ contains the Berry curvature contribution. We will review this later. Note that for the Rashba model without a magnetisation this contribution vanishes altogether. It is, however, in general nonzero.
Now the part coming from the band-off-diagonal matrix elements of the spin operator
\begin{equation}
	\displaystyle J^i_{j, 2} = \frac{1}{2} \, {\rm Tr} \, s_i^{od} \, \{ v_j^d, S_E \} + \frac{1}{2} \, {\rm Tr} \, s_i^{od} \, \{ v_j^{od}, S_E \}_{od} + {\rm Tr} \, s_i^{od} \, \{ v_j^{od}, n_E\}. 
\end{equation}
This is immediately recognised as the part of the conventional spin current that we usually evaluate for spin-1/2 electron systems, the part that gives $e/(8\pi)$ for the Rashba Hamiltonian. So we have to evaluate $J^i_{j, 1}$, $J^i_{j, 2}$, and $I^i_{j}$. The latter term is found from
\begin{equation}
		\displaystyle I^i_{j} \rightarrow   i \, {\rm tr} \, \int \frac{d^dq}{(2\pi)^{2d}} \, t_{i, {\bm q}} \bigg(\pd{S_{E{\bm q} {\bm Q}}}{Q_j}\bigg)_{{\bm Q} \rightarrow 0}.
\end{equation}
There is no band-diagonal part to this as $t_i$ is entirely band off-diagonal. The intrinsic contribution stemming from $S_E$ has already been covered in Ref. \onlinecite{liu2023}.

\subsection{First part of the conventional spin current for the extrinsic part of $S_E$}

We will begin by evaluating the extrinsic contributions from $S_E$. We first evaluate
\begin{equation}
	\displaystyle J^i_{j, 2} \rightarrow \frac{1}{2} \, {\rm Tr} \, s_i^{od} \, \{ v_j^d, S_E \}+ \frac{1}{2} \, {\rm Tr} \, s_i^{od} \, \{ v_j^{od}, S_E \}_{od}. 
\end{equation}
The diagonal velocity is ${\bm v}_d = (1/\hbar) (\partial H_0/\partial {\bm q})$. So this term takes the form
\begin{equation}\label{OD2}
\begin{aligned}
	\displaystyle J^i_{j, 2} =& \frac{1}{2} \sum_{mn} \, s^i_{mn} \, \{ v^j_d, S_E \}_{nm} + \frac{1}{2} \sum_{mn} \, s^i_{mn} \, \{ v^j_{od}, S_E \}_{nm}\\
	=& \frac{1}{2} \sum_{mn,m\neq n} \, \frac{1}{\hbar}\bigg( \pd{\varepsilon_m}{\bm q} + \pd{\varepsilon_n}{\bm q} \bigg)  s^i_{mn} \, S^E_{nm} - \frac{i}{\hbar}\, s^i_{mn}\{[\mathcal{R},H_0],S^E\}_{nm} . 	
\end{aligned}
\end{equation}

\subsection{The conserved spin current correction (torque dipole)}

The torque dipole term is
\begin{equation}
	\displaystyle I^i_{j} = i \, {\rm Tr} \,  t^i \bigg(\pd{S^E_{{\bm q} {\bm Q}}}{Q_j}\bigg)_{{\bm Q} \rightarrow 0}.
\end{equation}
In the eigenstate basis the off-diagonal contribution to this term is found from the equation
\begin{equation}\label{kQ}
	\arraycolsep 0.3ex
	\begin{array}{rl}
		\displaystyle \pd{S^E_{{\bm q}{\bm Q}}}{t} + \frac{i}{\hbar} \, [H_0, S^E_{{\bm q}{\bm Q}}] = & \displaystyle - \frac{i{\bm Q}}{2\hbar} \cdot \bigg\{\frac{DH_0}{D{\bm k}}, S^{E,0}_{\bm q} \bigg\}_{od}.
	\end{array}
\end{equation}
This expression determines the driving term on the RHS of Eq. \ref{kQ}. First of all, since the RHS of \ref{kQ} must be off-diagonal in the band index, and $S^E_{\bm q}$ is already off-diagonal in the band index, only the band-diagonal part of the velocity enters the driving term. This means the driving term is
\begin{equation}
	d^{\bm Q}_{mn} = - \frac{i{\bm Q}}{2\hbar} \cdot \bigg\{\frac{DH_0}{{D\bm q}}, S^{E,0}_{\bm q} \bigg\}_{mn} = - \frac{i{\bm Q}}{2 \hbar} \cdot \bigg( \pd{\varepsilon_m}{{\bm q}} + \pd{\varepsilon_n}{{\bm q}} \bigg) \, S^{E,0}_{{\bm q}, {mn}}- \frac{{\bm Q}}{2\hbar} \cdot \{[\mathcal{R},H_0], S^{E,0}_{\bm q} \}_{mn}.
\end{equation}

To determine the solution to \ref{kQ} we again refer to the interband coherence paper
\begin{equation}
	S^E_{{\bm qQ}, mn} = -\frac{i\hbar d^{\bm Q}_{mn}}{\varepsilon_m - \varepsilon_n}
\end{equation}
For our case this gives us
\begin{equation}\label{OD3}
	\arraycolsep 0.3ex
	\begin{array}{rl}
		\displaystyle S^E_{{\bm qQ}, mn} = & \displaystyle -\frac{{\bm Q}}{2} \cdot \bigg( \pd{\varepsilon_m}{{\bm q}} + \pd{\varepsilon_n}{{\bm q}} \bigg) \, \frac{S^{E,0}_{\bm q, {mn}} }{\varepsilon_m - \varepsilon_n} + \frac{i{\bm Q}}{2(\varepsilon_m - \varepsilon_n)} \{[\mathcal{R},H_0], S^{E,0}_{\bm q} \}_{mn}\\ [3ex]
		
		\displaystyle \pd{S^E_{{\bm q} {\bm Q}, mn}}{\bm Q} = & \displaystyle -\frac{1}{2} \bigg( \pd{\varepsilon_m}{{\bm q}} + \pd{\varepsilon_n}{{\bm q}} \bigg) \, \frac{S^{E,0}_{\bm q, {nm}} }{\varepsilon_m - \varepsilon_n} + \frac{i}{2(\varepsilon_m - \varepsilon_n)} \{[\mathcal{R},H_0], S^{E,0}_{\bm q} \}_{mn} \\ [3ex]
		
		\displaystyle I^i_{j} = & \displaystyle - \frac{i}{2} \sum_{mn} \,  t^i_{mn} \bigg( \pd{\varepsilon_m}{{\bm q}} + \pd{\varepsilon_n}{{\bm q}} \bigg) \, \frac{S^{E,0}_{\bm q, {nm}} }{\varepsilon_n - \varepsilon_m} - \frac{i\, t^i_{mn}}{2(\varepsilon_n - \varepsilon_m)} \{[\mathcal{R},H_0], S^{E,0}_{\bm q} \}_{nm}\\ [3ex]
		
		\displaystyle t^i_{mn} = & \displaystyle \frac{i}{\hbar} \, [H_0, s^i]_{mn} = \frac{i}{\hbar} (\varepsilon_m - \varepsilon_n) \, s^i_{mn} \\ [3ex]
		
		\displaystyle I^i_{j} = & \displaystyle -\frac{1}{2} \sum_{mn} \,\frac{1}{\hbar} s^i_{mn} \bigg( \pd{\varepsilon_m}{{\bm q}} + \pd{\varepsilon_n}{{\bm q}} \bigg) \, S^{E,0}_{\bm q, {nm}}+ \frac{i}{\hbar} \,s^i_{mn} \{[\mathcal{R},H_0], S^{E,0}_{\bm q} \}_{nm}.
	\end{array}
\end{equation}
which exactly cancels $J^i_{j, 2}$ in a non degenerate system.

\subsection{Other conventional current term}
We now evaluate
\begin{equation}
	\displaystyle J^i_{j, 1} = \frac{1}{2} \, {\rm Tr} \, s_i^d \, \{ v_j^{od}, S_E \}.
\end{equation}
In the eigenstate basis the inter-band velocity has the form $ {\bm v}_{od} = - \frac{i}{\hbar} [{\bf \mathcal R}, H_0]$, with matrix elements
\begin{equation}
	- i [{\bf \mathcal R}, H_0]_{mn} = - i {\bf \mathcal R}_{mn} (\varepsilon_n - \varepsilon_m).
\end{equation}
From the interband coherence paper, the inter-band part of the density matrix is
\begin{equation}
	S^E_{mn} = -i\hbar\frac{J_{mn}(n_E)}{(\varepsilon_m - \varepsilon_n)}.
\end{equation}
Then we have
\begin{equation}
	\displaystyle J^i_{j, 1} = -\frac{1}{2} \, \sum_{mn} s^i_{mm} \, \bigg\{  [{\bf \mathcal R}, H_0]_{mn}   \frac{J_{nm}(n_E)}{(\varepsilon_n - \varepsilon_m)}    +    \frac{J_{mn}(n_E)}{(\varepsilon_m - \varepsilon_n)} [{\bf \mathcal R}, H_0]_{nm} \bigg\}.  
\end{equation}
This can be simplified as
\begin{equation}
	\arraycolsep 0.3ex
	\begin{array}{rl}
	\displaystyle J^i_{j, 1} = & \displaystyle -\frac{1}{2} \, \sum_{mn} s^i_{mm} \, \bigg\{ {\bf \mathcal R}_{mn} (\varepsilon_n - \varepsilon_m)  \frac{J_{nm}(n_E)}{(\varepsilon_n - \varepsilon_m)}    +     \frac{J_{mn}(n_E)}{(\varepsilon_m - \varepsilon_n)}{\bf \mathcal R}_{nm} (\varepsilon_m - \varepsilon_n) \bigg\} \\ [3ex]
	
\displaystyle = & \displaystyle -\frac{1}{2} \, \sum_{mn} s^i_{mm} \, \bigg\{ {\bf \mathcal R}_{mn}  J_{nm}(n_E)   +     J_{mn}(n_E){\bf \mathcal R}_{nm} \bigg\} \\ [3ex]
	 
\displaystyle = & \displaystyle -\frac{1}{2} {\rm Tr}s_i\{\mathcal{R}_j,J_{od}(n_E)\}\,.
	 
	\end{array}
\end{equation}
This is the only contribution from the off-diagonal terms that needs to be considered. The cancellation of \ref{OD2} and \ref{OD3} is a general result for both intrinsic and extrinsic contributions

\subsection{The conventional spin current contributions from the band diagonal density matrix}
For this we only need calculate the remaining terms from IV. The first two conventional spin current terms are
\begin{equation}
\begin{aligned}
    \displaystyle J^i_{j, 1}  =& {\rm Tr} \, s_i^d \,  v_j^d n_E\\
    =&\frac{1}{\hbar}\sum_m s^i_{mm} n^E_m \pd{\varepsilon_m}{\boldsymbol{k}},
\end{aligned}
\end{equation}
and
\begin{equation}
\begin{aligned}
    \displaystyle J^i_{j, 2} =& \frac{1}{2} {\rm Tr} \, s_i^{od} \, \{ v_j^{od}, n_E\}\\
    =&-\frac{i}{2\hbar}\sum_{nm} s^i_{mn}\{[\mathcal{R}^j,H_0], n_E\}_{nm}\\
    =&-\frac{i}{2\hbar}\sum_{nm} s^i_{mn} \mathcal{R}^j_{nm} (n^E_m + n^E_n) (\varepsilon_m - \varepsilon_n). 
\end{aligned}
\end{equation}
So we now just have to evaluate $I^i_{j}$. Note although only $S_{E{\bm q} {\bm Q}}$ is present in the conserved spin current correction, there will still be a contribution from $n_E$ within it
\begin{equation}
		\displaystyle I^i_{j} \rightarrow   i \, {\rm Tr} \, \int \frac{d^dq}{(2\pi)^{2d}} \, t_{i, {\bm q}} \bigg(\pd{S_{E{\bm q} {\bm Q}}}{Q_j}\bigg)_{{\bm Q} \rightarrow 0}.
\end{equation}
This conserved contribution will be evaluated in two parts.

\subsection{$1^{st}$ part of the diagonal contribution to the torque dipole correction}

Recall the kinetic equation for $\rho$ to first order in $\bm Q$ is
\begin{equation}
    \displaystyle \pd{\rho_{{\bm q}{\bm Q}}}{t} + \frac{i}{\hbar} \, [H_{0{\bm q}}, \rho_{{\bm q}{\bm Q}}] + J(\rho_{{\bm q} {\bm Q}}) = \displaystyle - \frac{i{\bm Q}}{2\hbar} \cdot \bigg\{\frac{DH_{0{\bm q}}}{D{\bm q}}, \rho_{\bm q} \bigg\}.
\end{equation}
Focusing on the off-diagonal driving term containing $n_E$ we get
\begin{equation}
    \displaystyle \pd{\rho_{{\bm q}{\bm Q}}}{t} + \frac{i}{\hbar} \, [H_{0{\bm q}}, \rho_{{\bm q}{\bm Q}}] = \displaystyle - \frac{i{\bm Q}}{2} \cdot \bigg\{\boldsymbol{v}_{od}, n_{E\bm q} \bigg\}.
\end{equation}
The solution to the off-diagonal component will simply be
\begin{equation}
\begin{aligned}
    S_{\bm q Q}^{mn}=&-i\hbar\left(\frac{\left(-\frac{i\bm Q}{2}(\boldsymbol{v}_{mn}n_{E\bm q}^n+n_{E\bm q}^m\boldsymbol{v}_{mn})\right)}{\varepsilon_m-\varepsilon_n}\right)\\
    =&-\frac{i \bm Q}{2}\cdot\boldsymbol{\mathcal{R}}_{mn}(n_{E\bm q}^m+n_{E\bm q}^n)
\end{aligned}
\end{equation}
Now evaluating the term $I^i_{j}$ we find
\begin{equation}
\begin{aligned}
    I^i_{j,1}=&{\rm Tr}\,i\, t_{i, {\bm q}} \bigg(\pd{S_{E{\bm q} {\bm Q}}}{Q_j}\bigg)_{{\bm Q} \rightarrow 0}\\
    =&\frac{1}{2}\sum_{nm} t^i_{mn} \mathcal{R}_{nm}^j (n_{E\bm q}^m+n_{E\bm q}^n)\\
    =&\frac{i}{2\hbar}\sum_{nm} s^i_{mn} \mathcal{R}^j_{nm} (n^E_m + n^E_n) (\varepsilon_m - \varepsilon_n).
\end{aligned}
\end{equation}
This cancels exactly with $J^i_{j, 2}$.

\subsection{$2^{nd}$ part of the diagonal contribution to the conserved spin current correction}
This final contribution can't be explicitly calculated so I will just outline how it is to be calculated. First we start kinetic equation of $\rho$ to first order in $\bm Q$
\begin{equation}
    \displaystyle \pd{\rho_{{\bm q}{\bm Q}}}{t} + \frac{i}{\hbar} \, [H_{0{\bm q}}, \rho_{{\bm q}{\bm Q}}] + J(\rho_{{\bm q} {\bm Q}}) = \displaystyle - \frac{i{\bm Q}}{2\hbar} \cdot \bigg\{\frac{DH_{0{\bm q}}}{D{\bm q}}, \rho_{\bm q} \bigg\}.
\end{equation}
Now, we care about the band-diagonal driving term and $n_{{\bm q}{\bm Q}}$
\begin{equation}
    \displaystyle \pd{n_{{\bm q}{\bm Q}}}{t} + J(n_{{\bm q} {\bm Q}}) = \displaystyle - \frac{i{\bm Q}}{2\hbar} \cdot \bigg\{\pd{H_{0{\bm q}}}{{\bm q}}, n_{\bm q} \bigg\} - \frac{{\bm Q}}{2\hbar} \cdot \{[\mathcal{R},H_0], S^{E,0}_{\bm q} \}_{d}.
\end{equation}
This equation must be solved for $n_{{\bm q}{\bm Q}}$, so we can get our last off-diagonal component to $S_{\bm q Q}^{mn}$
\begin{equation}
    S_{\bm q Q}^{mn} = -i\hbar\frac{J_{mn}(n_{{\bm q}{\bm Q}})}{(\varepsilon_m - \varepsilon_n)}.
\end{equation}
Once this has been found, you simply take the trace of
\begin{equation}
    I^i_{j,2}={\rm Tr}\,t_{i, {\bm q}} \bigg(\pd{S_{{\bm q} {\bm Q}}}{Q_j}\bigg)_{{\bm Q} \rightarrow 0}.
\end{equation}

\section{Total spin current}
Writing out all the surviving extrinsic contributions to the spin current we have
\begin{equation}
    \mathcal{J}^i_j = -\frac{1}{2}\, {\rm Tr}\,[s_i]_d\{\mathcal{R}_j,J_{od}(n_E)\} + \frac{1}{\hbar} \sum_m s^i_{mm} n^E_m \pd{\varepsilon_m}{\boldsymbol{k}}+I^i_{j,2}\,.
\end{equation}

\end{document}